# Towards Chirality Control of Graphene Nanoribbons Embedded in Hexagonal Boron Nitride

Hui Shan Wang[1,2,6†], Lingxiu Chen[1,6†], Kenan Elibol[3†#], Li He[4†], Haomin Wang[1,2,6*], Chen Chen[1,2,6], Chengxin Jiang[1,5,6], Chen Li[7$], Tianru Wu[1,6], Chun Xiao Cong[8], Timothy J. Pennycook[3$], Giacomo Argentero[3], Daoli Zhang[4], Kenji Watanabe[9], Takashi Taniguchi[9], Wenya Wei[10,11], Qinghong Yuan[10,11], Jannik C. Meyer[3&*], Xiaoming Xie[1,2,5,6]

[1] State Key Laboratory of Functional Materials for Informatics, Shanghai Institute of Microsystem and Information Technology, Chinese Academy of Sciences, 865 Changning Road, Shanghai 200050, P. R. China.

[2] Center of Materials Science and Optoelectronics Engineering, University of Chinese Academy of Sciences, Beijing 100049, P. R. China.

[3] Faculty of Physics, University of Vienna, Boltzmanngasse 5, Vienna 1090, Austria.

[4] School of Optical and Electronic Information, Huazhong University of Science and Technology, Wuhan 430074, P. R. China.

[5] School of Physical Science and Technology, ShanghaiTech University, Shanghai 201210, P. R. China.

[6] CAS Center for Excellence in Superconducting Electronics (CENSE), Shanghai 200050, P. R. China.

[7] Department of Lithospheric Research, University of Vienna, Althanstrasse 14, 1090 Vienna, Austria.

[8] State Key Laboratory of ASIC & System, School of Information Science and Technology, Fudan University, Shanghai 200433, P. R. China.

[9] National Institute for Materials Science, 1-1 Namiki, Tsukuba 305-0044, Japan.

[10] State Key Laboratory of Precision Spectroscopy, School of Physics and Material Science, East China Normal University, Shanghai 200062, P. R. China.

[11] Centre for Theoretical and Computational Molecular Science, Australian Institute for Bioengineering and Nanotechnology, The University of Queensland, Brisbane, QLD 4072, Australia

† These authors contribute equally to this work.

# Present address: now at School of Chemistry, Trinity College Dublin, CRANN - Advanced Microscopy Laboratory, Unit 27/29 Trinity Enterprise Centre，Pearse St, Dublin 2, Ireland.

$ Present address: now at Electron Microscopy for Materials Research (EMAT), University Antwerpen, Groenenborgerlaan 171, 2020 Antwerpen, Belgium.

& Present address: now at Institute for Applied Physics and Natural and Medical Sciences Institute, University of Tübingen, Tübingen, Germany.

* Electronic mail: hmwang@mail.sim.ac.cn, jannik.meyer@uni-tuebingen.de

**The integrated in-plane growth of two dimensional materials (*e.g.* graphene and hexagonal boron nitride (*h*-BN)) with similar lattices, but distinct electrical properties, could provide a promising route to achieve integrated circuitry of atomic thickness. However, fabrication of edge-specific graphene nanoribbons (GNR) in the lattice of *h*-BN still remains an enormous challenge for present approaches. Here we developed a two-step growth method and successfully achieved sub-5 nm-wide zigzag and armchair GNRs embedded in *h*-BN, respectively. Further transport measurements reveal that the sub-7 nm-wide zigzag GNRs exhibit openings of the band gap inversely proportional to their width, while narrow armchair GNRs exhibit some fluctuation in the bandgap-width relationship. Transistors made of these GNRs with large bandgaps (>0.4 eV) exhibit excellent electronic performance even at room temperature (*e.g.* conductance on-off ratio more than $10^5$ and carrier mobility more than 1,500 $cm^2\ V^{-1}\ s^{-1}$). An obvious conductance peak is observed in the transfer curves of 8-10 nm-wide zigzag GNRs while it is absent in most of armchair GNRs of similar width. Magneto-transport experiments show that zigzag GNRs exhibit relatively small magneto-conductance (MC) while armchair GNRs have much higher MC than zigzag GNRs. This integrated lateral growth of edge-specific GNRs in *h*-BN brings semiconducting building blocks to atomically thin layer, and will provide** a promising route to achieve intricate nanoscale electrical circuits on high-quality insulating *h*-BN substrates.

Graphene nanoribbons (GNRs), a quasi-one-dimensional graphene nanostructure, can exhibit either quasi-metallic or semiconducting behavior, depending on its specific chirality, including width, lattice orientation, and edge structure [1]. The unique properties of GNR make it a promising substitute to engineer prospective nano-electronics. There are two groups of GNRs, that differ by the edge type and are called zigzag (ZZ) and armchair (AC) [2]. Recent experiments [3,4] show that zigzag graphene nanoribbons (ZGNRs) potentially offer exotic electronic properties, such as ferromagnetism and half metallicity. Theory [5] predicts that all armchair graphene nanoribbons (AGNRs) are semiconducting with a band gap that is inversely proportional to their width. Normally, both ZGNRs and AGNRs have distinct electronic states and scattering properties [6,7] as well as unique chemical properties [8,9]. Practically, their edges are prone to intrinsic and extrinsic modifications [10-12]. Recent hetero-integration of graphene with *h*-BN by both van der Waals stacking [13,14] and in-plane covalent bonding [15-19] has exhibited an advantage in high chemical/mechanical stability, and may lead to emergent electronic properties that are fundamentally distinct from those of the constituents. However, such a hetero-structure with chirality-controllable GNRs is still difficult to achieve.

In this work, we report the successful control over the chirality of mono-layer GNRs directly embedded in $h$-BN nano-trenches, whose direction can be modulated by different catalytic cutting particles. Both ZGNRs and AGNRs narrower than 5 nm can be fabricated. Scanning transmission electron microscopy investigation shows that in-plane epitaxy was realized at the boundary of graphene and $h$-BN with controlled chirality at the edge along the GNR while developed laterally. Further electrical investigation reveals that all narrow ZGNRs exhibit a bandgap larger than 0.4 eV while narrow AGNRs exhibit a relatively large variation in band-gap. Transistors made of GNRs with large bandgaps exhibit on-off ratios of more than $10^5$ at room temperature with carrier mobilities higher than 1,500 cm$^2$ V$^{-1}$ s$^{-1}$. With such a method, narrow GNRs along high symmetry lattice-orientations and with smooth edges in $h$-BN provide exciting possibilities for designer electronics and fundamental research in condensed matter physics.

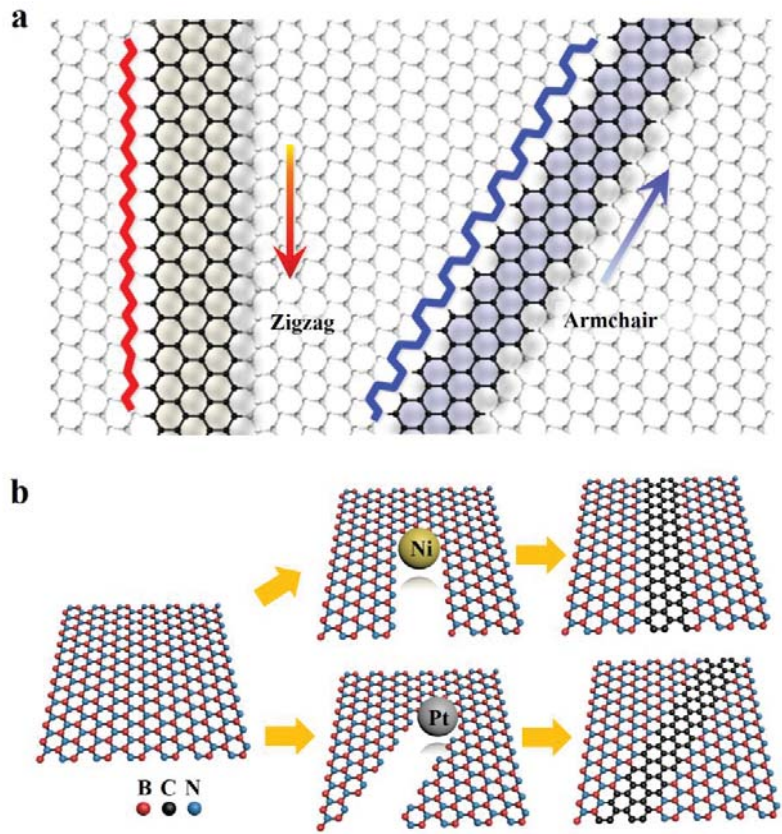

**Fig. 1 | Synthetic strategy to oriented GNRs embedded in *h*-BN. a** Sketch of ZGNR and AGNR embedded in *h*-BN. **b** Synthetic strategy to GNRs with crystallographic edge orientations. Different particles cut nano-trenches with different edge topologies in an *h*-BN layer. The edge-specific nano-trenches are used to define the growth and the dimensions of GNRs along the ZZ or AC direction, respectively.

**Fig.** 1a depicts the structure of ZGNR and AGNR embedded in an *h*-BN lattice layer. Recent theoretical work predicts that ZGNRs embedded in *h*-BN exhibits strong current-polarization because of their asymmetrical edges [20], while all the AGNRs are semiconductors [21]. There have been several attempts to grow graphene ribbons via plasma enhanced chemical vapor deposition (CVD) [22] or thermal CVD [23]. Because of the difficulties in fabrication, little experimental work on chirality-controllable GNRs embedded in *h*-BN has been reported.

**Fig.** 1b schematically illustrates the synthesis of oriented GNRs in *h*-BN via a two-step method (crystallographically cutting and template growth). A single-crystalline *h*-BN layer exhibits a perfect crystallinity with an atomically flat surface. The nano-trenches in *h*-BN were obtained by nanoparticle-catalyzed cutting, which exhibits obvious orientation selectivity and particle dependence. In our experiments, nickel (Ni) particles lead to nano-trenches along the ZZ direction while platinum (Pt) particles produce AC-oriented nano-trenches in the *h*-BN layer (**Fig.** S1). Subsequently, the trenches were filled with GNRs by gaseous catalyzed CVD [24] (**Fig.** S2). Lateral connectivity between GNRs and *h*-BN makes in-plane hetero-structures with lattice coherence. Finally, ZGNRs or AGNRs embedded in a continuous *h*-BN sheet were successfully fabricated using *h*-BN substrates. First principles DFT calculations were carried out to help understand the experimental observations of GNR growth (**Fig.** S5-8). Considering particle-catalyzed etching as the opposite reaction of the CVD growth process, particle-catalyzed growth of GNR may also occur during the etching of *h*-BN under certain circumstances.

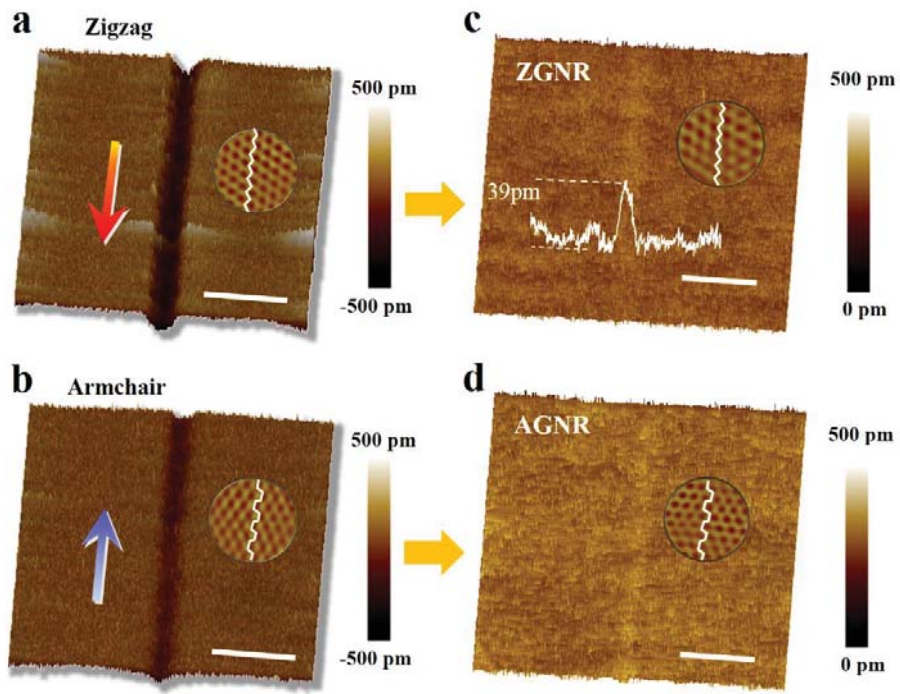

**Fig. 2 | Edge-specific nano-trenches and GNRs embedded in top layer of *h*-BN.** 3D AFM height images of sub-5 nm ultra-narrow **a** ZZ-oriented and **b** AC-oriented nano-trenches taken after crystallographically selective nanoparticle etching; 3D AFM height image of sub-5 nm ultra-narrow **c** ZGNR and **d** AGNR taken after graphene growth. The circular insets show atomic-resolution friction images of the *h*-BN and show the crystallographic orientation of nano-trenchs and GNRs. Scale bars: 20 nm.

**Fig.** 2 gives examples of edge-specific nano-trenches and ultra-narrow GNRs embedded in the top layer of *h*-BN substrates. The bright protrusions near the edges in **Fig.** 2 are due to a known imaging instability [25], rather than edge defects or impurities. **Fig.** 2a shows a 3D AFM height image of a sub-5 nm ZZ-oriented nano-trench obtained via Ni nanoparticle-assisted etching. Upon optimizing the AFM imaging, we found the edges defined by nano-particle-catalyzed etching technique to be straight and close to atomically smooth. **Fig.** 2c presents a typical height image of a ZGNR grown in an *h*-BN nano-trench. The height profile exhibits a ~39 pm out-of-plane distortion on the surface of the *h*-BN. It is most likely due to lattice distortions caused by mismatched lattice constants and coefficients of thermal expansion between graphene and *h*-BN. The tiny out-of-plane distortions exclude the possibility of the formation of multilayered GNRs. Similarly, **Fig.** 2b and 2d give the AFM height images of an ultra-narrow AC nano-trench obtained via Pt particle-assisted etching and an ultra-narrow AGNR, respectively. The corresponding AFM friction images of the ZGNR and AGNR are given in Supplementary **Fig.** S9. The circle inset shows the atomic-resolution friction image of *h*-BN, which confirms that the crystallographic orientation of the trench/GNR is along the zigzag direction in **Fig.** 2a & 2c and the armchair lattice direction in **Fig.** 2b & 2d.

We also tried to image the step-edges of *h*-BN trenches. Atomic resolution analysis on the *h*-BN edge of zigzag oriented trench and armchair oriented trench are given in **Fig.** S10 and S11, respectively. As shown in these figures, both edges are atomically smooth.

**Fig.** S12a–d shows examples of ZZ-oriented nano-trenches in *h*-BN substrates. As shown in **Fig.** S12a, the nano-trenches mutually exhibit separation angles of 60°, indicating anisotropic etching along a specific crystallographic direction. The inset of **Fig.** S12a confirms that the crystallographic orientations of the *h*-BN trenches/edges follow the ZZ pattern. This indicates that the straight nano-trenches along ZZ direction of the *h*-BN sheet are driven by the energetically favored Ni particle-zigzag BN interface. Similar results are shown in **Fig.** S3a and S3b. The width distributions of ZZ-oriented trenches in our optimized etching condition are shown in **Fig.** S4a.

Both the width and length of the trenches depend on the process parameters. Nanometre-sized trenches, narrower than 5 nm in width, can be reproduced via optimizing the process (see Supplementary **Table** S1). It might be argued that relatively "blunt" AFM tips would not resolve the exact depth of the narrow trenches and their abrupt height change, but these trenches can be reasonably deduced to be monolayer by measuring wider nano-trenches.

Wider nano-trenches (**Fig.** S12d) could be obtained by extending the duration of etching or increasing the temperature (see Supplementary **Table** S1). The ~78 nm-wide trench in **Fig.** S12d had a depth of ~0.340 nm with a bottom roughness comparable to the surface of *h*-BN. The results indicate that the wider trench is mono-layered. The extremely anisotropic two-dimensional (2D) nature of *h*-BN predominantly confined the etching to the top single-atomic layer. Note that the AFM tip size always results in over-estimation of the width of the nano-trenches.

**Fig.** S12e–h presents typical AFM height images of narrow ZGNRs embedded in the nano-trenches of the *h*-BN substrate. Obviously, the GNRs inherit their alignment nature from the templates of the ZZ nano-trenches. For narrow GNRs, out-of-plane height in the range of 20–40 pm can be seen on the surface of the *h*-BN.

To study the crystallinity of the embedded ZGNRs, we investigate here freely suspended thin *h*-BN flakes embedded with ZGNRs via transmission electron microscopy (TEM) and scanning transmission electron microscopy (STEM). Low electron energies (80 keV for TEM and 60 keV for STEM) were used to minimize radiation damage [ 26 ]. Details about the fabrication of the TEM/STEM samples are given in the Methods section and **Fig.** S13. The sample was imaged via TEM in advance. As shown in **Fig.** S13, a TEM dark-field (DF) image clearly shows the presence of graphene in the freely suspended thin *h*-BN flake. Further investigation on this sample was performed in a Nion UltraSTEM 100 operated at 60 kV. **Fig.** S14 shows a medium-angle annular dark-field (MAADF) image of a small portion of the suspended heterostructure using electron scattering angles between 60 and 200 mrad and a high-angle annular dark-field (HAADF) image (ca. 80−240 mrad), respectively. For overview images (**Fig.** S12i), we found that the contrast of adsorbed contamination could be almost completely eliminated by calculating a difference MAADF-α×HAADF, where α is adjusted to minimize the visibility of contamination, and thereby reveal the ribbons or other features of interest already at low magnification.

**Fig.** S12i is such a MAADF-α×HAADF image showing an area with an embedded ZGNR. The bright white particles observed are primarily silicon oxide. The white dashed box in **Fig.** S12i shows the position of a ZGNR (a dark line). The MAADF image of **Fig.** S12j shows a zoomed-in view of the ZGNR field in **Fig.** S12i. As shown in **Fig.** S12j, the ZGNR is straight and uniform at the nanometer-scale, and the measured width is ~3.2 nm (within the accuracy of the STEM measurements). The *h*-BN-GNR boundaries are sharp and highly crystalline. The length of the GNR observed in **Fig.** S12j is more than 80 nm. The seamless construction of the GNR is mainly attributed to the gaseous catalytic growth [24]. **Fig.** S12k shows a Wiener-filtered [27] MAADF image of the region inside the dashed frame in **Fig.** S12j. The GNR is clearly along the zigzag direction of the *h*-BN lattice, indicating that the embedded GNR is ZZ lattice oriented. It is also clear that the whole lattice is connected in a continuous hexagonal network. Due to contamination from hydrocarbon, it is difficult to distinguish carbon, boron and nitrogen atoms via electron energy loss spectroscopy (EELS) mapping. However, the boundaries between GNR and *h*-BN still can be distinguished straightforwardly in **Fig.** S12i and S12j. In this area, the *h*-BN flake has a thickness of about 3-5 layers (see Supplementary **Fig.** S13), and the GNRs exhibit obvious contrast to the surrounding *h*-BN region. From the atomic resolution image in **Fig.** S12k, it is clear that the ZGNR is in lattice coherence with the *h*-BN.

It is necessary to mention that it is extremely challenging to obtain completely clean surfaces in the GNR areas even after multiple-round annealing treatments. Practically, contamination like hydrocarbon impurities may be directly absorbed from the atmosphere onto the sample surface. The contamination can move and accumulate at energetically favored areas (e.g. wrinkles, grain boundaries and defects) during the annealing treatments. In addition, mobile contamination may be pinned into place by the electron beam.

The contamination makes atomic-level characterization of GNR-BN interface by electron microscopy and EELS very difficult.

**Fig.** S15a–d shows AC-oriented nano-trenches obtained in *h*-BN substrates. As shown in **Fig.** S15a, the nano-trenches exhibited mutual separation angles of 60° with no obvious correlation to the direction of the gas flow. Separation angles of odd multiples of 30° were occasionally observed. The white hexagons in the inset of **Fig.** S15a help in identifying that the *h*-BN trenches/edges are primarily along AC-direction. This indicates that the crystallographically selective chemical reaction between Pt particles and *h*-BN has a lower activation energy along the AC-direction in *h*-BN. The nano-trenches shown in **Fig.** S15a–c are AC-oriented trenches narrower than 5 nm in width. Wider nano-trenches can be obtained by increasing the etching duration (**Table** S2). In **Fig.** S15d, it is also found that the ~23 nm-wide trench is mono-layered because it has a depth of ~0.334 nm with a smooth bottom of *h*-BN. Similar to the ZZ trenches, these narrow AC trenches were also reasonably judged to be mono-layered.

**Fig.** S15i-j show MAADF images of a small portion of the suspended *h*-BN layers embedded with armchair GNRs. The white arrow in **Fig.** S15i points to a GNR oriented along the AC direction. To the lower left of the AGNR appear

two parallel ZGNRs roughly perpendicular to the AC one. Note that the suspended *h*-BN flake buckled a little bit in the field of view. In **Fig.** S15j, we see a close-up view of the AGNR embedded in the few-layered *h*-BN sheet. The AGNR is straight and uniform in nanometer-scale resolution, and has a width of ~5.3 nm (within the accuracy of our STEM measurements). The *h*-BN-GNR hetero-structure is highly crystalline and has lattice coherence. **Fig.** S15k is the Fourier transform of the image shown in **Fig.** S15j, from which the lattice orientation of the armchair GNR is determined. Similar results are shown in **Fig.** S3c and S3d. The width distributions of AC-oriented trenches in our optimized etching condition are shown in **Fig.** S4b.

The nanoparticle catalytic etching plays very important roles in determining the orientation and width of nano-trenches, and finally those of GNRs in *h*-BN. The mechanism behind this is worth discussing. Our experiments show that the anisotropic cutting and etching of *h*-BN sheets depend on the type of catalytic nanoparticle and etching agent ($H_2$ gas). Obviously, the appearance of long and straight ZZ and AC nano-trenches indicates that there are two energetically favored directions during the nanoparticle catalytic cutting on *h*-BN. The catalyst-dependent directional cutting behavior (nickel for ZZ trenches and platinum for AC trenches) shown in this experiment indicates that the interaction at the metal-*h*-BN interface is thermodynamically more stable than the H-terminated *h*-BN edge in terms of formation energy.

Due to the binary composition of *h*-BN, the edge configurations of *h*-BN trenches can be either B-rich or N-rich for ZZ edges, or B-N for AC edges, leading to a complicated edge chemistry and metal-BN interfaces. Experimentally, the Ni−$ZZ_B$ interface should have similar thermodynamic stability as the Ni−$ZZ_N$ interface, and both of them are more energetically favorable than the Ni-$AC_{BN}$ interface in optimized experimental conditions. The Pt-$AC_{BN}$ interface is slightly more stable than the Pt−$ZZ_B$ and Pt−$ZZ_N$ interfaces at high temperature and low $H_2$ gas pressure. In addition, the presence of $H_2$ gas is essential for catalytic etching of *h*-BN as $H_2$ can continuously convert the etched B and N atoms from the *h*-BN lattice into $BH_x$/$NH_x$/$BNH_x$.

In these experiments, it is found that the narrower the nano-trench is, the straighter it is. The ultra-narrow trenches are normally created by very small particles produced by extending the annealing duration before the *h*-BN substrates were heated to etching temperature. Recent theoretical investigation [28] reveals that a smaller metal nanoparticle would maximize its contact with either AC or ZZ edges and then move along the more stable metal–AC or metal–ZZ interfaces. This indicates that smaller nanoparticles would be better than larger ones for maintaining a straight cut.

GNRs grew via a step-flow mechanism from two step-edges of the *h*-BN top-layer trench [29]. Orientation of the graphene edge can be modified via tuning ethyne and silane flow in the CVD process [23]. GNRs grew along the atomic step-edge of the top layer on the *h*-BN and developed laterally. Finally, the GNRs developed from opposite sides of the trenches coalesced into a complete GNR. Atomic-resolution STEM images show that the GNRs are highly crystalline and have a lattice coherence with the top layer of *h*-BN.

The high-resolution noncontact atomic force microscopy (nc-AFM) measurement was carried out to investigate the boundaries of GNR-*h*-BN. The results are given in **Fig.** S16 and S17. The high-resolution AFM investigation confirm that both ZGNR embedded in *h*-BN and AGNR embedded in *h*-BN indeed have atomically smooth edges within tens of nanometers.

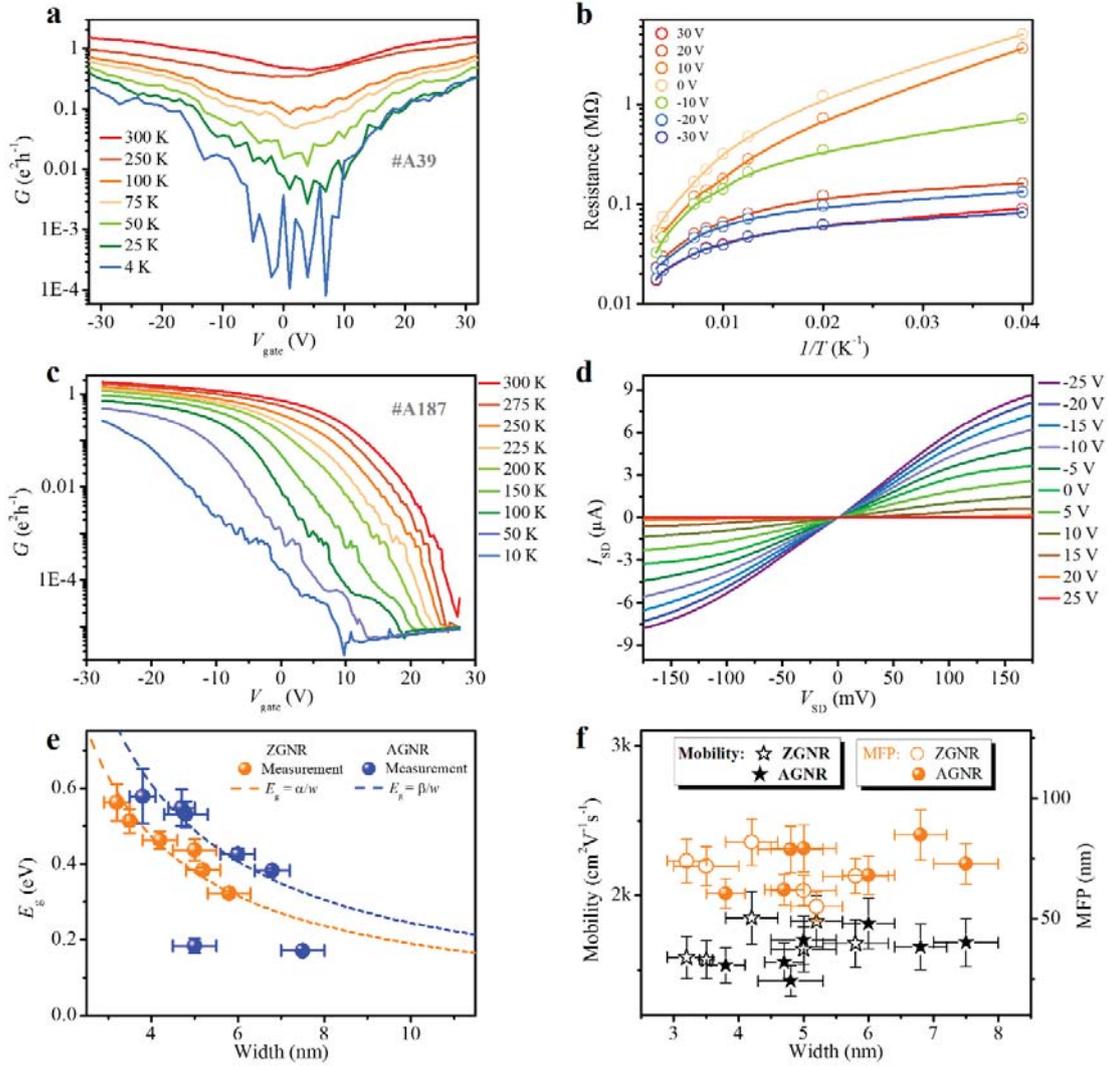

**Fig. 3 | Electronic transport through GNR devices on *h*-BN. a** Conductance (*G*) of an AGNR device (sample #A39) with a width of ~5 nm as a function of the back gate voltage ($V_{gate}$) at different temperatures. **b** Arrhenius plot of resistance of the AGNR FET under different $V_{gate}$ values at temperatures from 25 to 300 K. The solid curves are fits based on a simple two-band (STB) model. **c** *G* versus $V_{gate}$ for sample #A187 with a width of ~4.8 nm. The AGNR FET can be completely switched off even at room temperature. **d** $I_{SD} - V_{SD}$ characteristics recorded under $V_{gate}$ from -25 to 25 V for the device in **c** at 300K. **e** Band gap ($E_g$) extracted from experimental data for GNRs *versus* their ribbon width (*w*). Error bars represent standard deviation of uncertainty in width measurement and gap extraction. Some widths measured via TEM exhibits less uncertainty than those measured via AFM. The orange dashed line is a fit of the ZGNR data into an empirical formula of $E_g$ (eV) = 1.89/*w* where *w* is in unit of nm. The blue dashed line is a fit of the AGNR data into an empirical formula of $E_g$ (eV) = 2.44/*w*. **f** Mobility and mean free path (MFP) extracted from the GNR devices. It is noted that most narrow GNRs show carrier mobility higher than 1,500 $cm^2V^{-1}s^{-1}$, and their scattering MFP was estimated to be more than 50 nm.

Field-effect transistors (FET) were fabricated to investigate the electrical properties of the GNRs. A ~5 nm-wide AGNR (**Fig.** S20) was fabricated into a FET with a channel length ($L$) of ~278 nm (sample #A39). Its conductance $G$ versus $V_{gate}$ at different temperatures is shown in **Fig.** 3a, and exhibits obvious modulations with respect to back-gate voltage ($V_{gate}$). Its field-effect carrier mobility $\mu = \frac{\frac{dG}{dV_{gate}} \cdot L}{C_{gs}}$ at 300 K is about 1,700 cm$^2$V$^{-1}$s$^{-1}$, where the effective capacitance $C_{gs}$ = ~3.71 pF m$^{-1}$. **Fig.** 3b shows the temperature dependence of the resistance under different $V_{gate}$. The band gap ($E_g$) extracted by the fitting resistance-temperature curve is 183 ± 19 meV according to the simple two-band (STB) model (Please see the details of the STB model in the Supplementary Text), where both thermal activation and contact resistance have been taken into account. Another sub-5 nm-wide AGNR FET (**Fig.** S21) measured exhibits a relatively high on-off ratio in conductance. As shown in **Fig.** 3c, the AGNR device (sample #A187) in a channel length of ~236 nm exhibited $G_{on}/G_{off} > 10^5$ at 300 K. **Fig.** 3d shows the $I_{SD}$-$V_{SD}$ characteristics of the AGNR device recorded at $T$ = 300 K under $V_{gate}$ from -25 to 25 V (see $I_{SD}$-$V_{SD}$ curves at $T$ = 100 K in **Fig.** S18a). The $I_{SD}$-$V_{SD}$ curves under all $V_{gate}$ are almost linear before $I_{SD}$ saturates both in 300 K and 100 K. This indicates that the Pd-GNR contact resistance is small or absent, and the main reason may be that Pd has a high work function and good wetting interactions with GNRs

[30,31]. The extracted band gap value using the STB model is ~531 meV, and the field effect mobility is ~1,428 cm$^2$V$^{-1}$s$^{-1}$. The scattering mean free path (MFP) was also estimated to be ~79 nm as the relatively high $G = 1.8\frac{e^2}{h}$ at 300 K when $V_{gate}$ = -25 V.

Similarly, the electronic transport properties of a typical sub-5 nm-wide ZGNR FET (sample #Z143) were measured (**Fig.** S19), and their band-gap/carrier mobility/MFP are also extracted. Obviously, the narrow ZGNRs exhibit a band gap > 0.4 eV. The band gap extracted as a function of the corresponding ribbon width for all narrow GNRs is plotted in **Fig.** 3e. As shown in **Fig.** 3e, it is noted that the band gap scaled inversely with ribbon width both for ZGNRs and AGNRs. The width dependence of the band gap for ZGNRs fits well with the function $E_g$ (eV) ~ $\alpha/w$ ($w$ is in units of nm), where parameter $\alpha \approx$ 1.89 eV nm. Similarly, the fitting parameter $\beta \approx$ 2.44 eV nm for AGNRs. All of them display obvious semiconducting characteristics. For the narrow ZGNRs, the narrower the ribbon, the higher the band gap. The gap opening in the ZGNRs embedded in *h*-BN is attributed to the combined influence from e–e interactions [3], uniaxial strain [32] and the stacking order on *h*-BN [ 33 ]. For the narrow AGNR, the band gap does not scale monotonically with ribbon width, and there is some fluctuation on the band-gap scaling. According to first-principles DFT calculations [6], there are three groups of AGNRs: with $3n$, $3n+1$ and $3n+2$ (for $n = 1, 2, 3, …$) rows of carbon atoms across their width. AGNRs in the $3n+2$ group normally exhibit much smaller band gaps than those in the other two groups. Armchair ribbons occasionally observed with relatively small band gaps probably belong to the $3n+2$ group. Consequently, the experimental results are in agreement with the predicted mechanism of bandgap opening caused by the lateral confinement of the charge carriers, and indicate that the AGNRs are comprised of very long precise edge segments.

The carrier mobility and MFP of the GNRs are summarized in **Fig.** 3f. The mobility of the narrow GNRs is all at ~1,500 $cm^2V^{-1}s^{-1}$ and the scattering MFPs were estimated to be more than 50 nm. The results directly verify that carriers are not scattered in more than 50 nm long-distance of both ZGNRs and AGNRs. This indicates that the segments with precise edges in the narrow GNRs are more than 50 nm. The transport results together with the STEM data convinced us that edge-specific GNRs embedded in *h*-BN have been successfully fabricated in *h*-BN substrates.

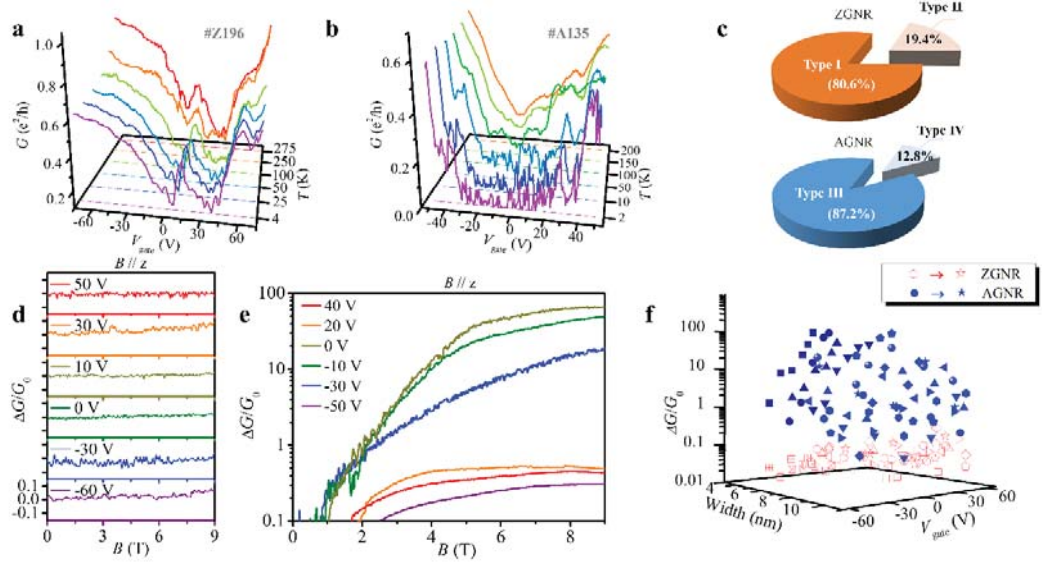

**Fig. 4 | Field effect and magneto-electrical properties in relatively wide GNRs in *h*-BN.** The typical transfer curves of **a** a ~8.9 nm-wide ZGNR sample (#Z196) and **b** a ~9.5 nm-wide AGNR sample (#A135) at different temperatures. **c** Pie charts of the type distribution in edge-specific GNRs according to their characteristic of conductance peak. In the pie chart of ZGNR, Type 'I' indicates ZGNRs whose conductance peaks survive even at high temperature (a typical example: #Z196), and Type 'II' represents ZGNRs whose conductance peaks disappear at high temperature; In pie chart of AGNRs, Type 'III' is AGNRs whose transfer curves are in absence of the conductance peaks (a typical example: #A135), and Type 'IV' is AGNRs whose transfer curves are of a tiny peak at low temperature. **d** Normalized magnetic conductance (MC) of sample #Z196 obtained at different $V_{gate}$ at 4 K, the magnetic field was applied perpendicular to substrate plane. **e** Normalized MC of sample #A135 measured at different $V_{gate}$ at 2 K with the magnetic field perpendicular to substrate. **f** Distribution of normalized MC of all GNRs measured as a function of $V_{gate}$ for GNRs of different widths. All MCs are measured in a magnetic field of 9 T perpendicular to substrate. The open symbols (red) indicate ZGNRs and the solid symbols (blue) represent AGNRs.

The field effect properties of the edge-specific GNRs of widths of 8 to 10 nm were systematically investigated. **Fig.** 4a shows $G$ *vs* $V_{gate}$ at different temperatures for a typical ZGNR device (sample #Z196) with a channel length of ~190 nm and a width of ~8.9 nm. A conductance peak ($V_{gate}$ = ~15 V) is observed at 4 K. The peak remains almost unchanged in conductance even when the temperature increases to 275 K. This weak metallic behavior of conductance indicates that the peak position may exhibit electronic states. **Fig.** 4b shows the relationship of conductance as a function of $V_{\text{gate}}$ with temperature for a typical AGNR FET (sample #A135) with a channel length of ~261 nm and a width of ~9.5 nm. Unlike the ZGNR, the AGNR device does not exhibit an obvious conductance peak near its charge neutrality point. As shown in **Fig.** 4b, its conductance exhibited a drastic V-/U-shape drop with respect to $V_{\text{gate}}$. This phenomenon suggests a band-gap opening in the nanoribbon.

In total, we measured the electronic properties of 36 ZGNRs and 39 AGNRs whose width ranges from 8 to 10 nm. The GNR devices can be classified into four types according to their electronic transport characteristics (see details in **Fig.** S22-24). The statistics of GNRs of different types are shown in **Fig.** 4c. It is found that all ZGNRs exhibit a clear peak in their transfer curves, while ~87.2% of AGNRs are absent of the peaks in their transfer curves. The conductance peak of ~80.6% ZGNRs can survive at 275 K. The conductance peak is believed to be related to the density of states at the zigzag edges of graphene [34].

Magneto-transport properties of the relatively wide GNRs were also investigated. **Fig.** 4d shows the relationship of normalized magnetic conductance (MC) *vs B* with $V_{gate}$ in a ZGNR (sample #Z196). The MC curves are normalized by using $\Delta G/G_0 = [G(B)-G_0]/G_0$, where $G(B)$ represents conductance measured at a magnetic field of $B$ and $G_0$ is the conductance measured in the absence of a magnetic field. Only a magnetic response of less than 10% is observed at some certain $V_{gate}$. Similar measurements were performed with an AGNR (sample #A135). As shown in **Fig.** 4e, a large positive MC of 7,000% ($V_{gate}$ = 0 V) is found at 2 K. The values of the MC are positive at all $V_{gate}$ and can be readily tuned by $V_{gate}$. It is also found that there is a saturated trend for MC at all $V_{gate}$ in **Fig.** 4e (more MC results are shown in **Fig.** S25-30). The normalized MCs of all GNRs obtained under 9 T at different $V_{gate}$ are plotted in **Fig.** 4f. As shown in **Fig.** 4f, MCs of all the ZGNRs are positive, but less than 10%, while almost all of the AGNRs exhibits a positive MC greatly higher than 10%.

The experimental results clearly show that the ZGNRs exhibit relatively small MC while AGNRs have larger MC. It is believed that the mechanism for the MC is related to their chirality. Further experiments and theoretical studies will be carried out to understand the mechanism responsible for the MC behavior in the edge-specific GNRs.

This work reports a strategic process for the fabrication of specific GNRs in the top atomic layer of insulating $h$-BN. The process combines traditional top-down (crystallographic cutting of an $h$-BN layer) and bottom-up approaches (GNR template growth) together. The process allows for the growth of electrically isolated components (graphene nanoribbons) in continuous two-dimensional insulating $h$-BN sheets with seamless in-plane heterojunctions ensuring that the components retain distinct electronic properties. The components do not alter the nature of the $h$-BN base in the mechanical strength, flexibility and optical transparency, which is desired in flexible, transparent electronics. Electrical transport measurement shows that ZGNRs and AGNRs exhibit obvious differences in their transfer curves and magneto-conductance. The integrated growth of semiconducting GNRs on high-quality insulating substrates provides the fundamental building blocks of atomically thin ULSI (ultra-large-scale integration) circuitry. Beyond fabrication, ZGNRs in contact with a superconductor are predicted to exhibit a topological electronic state that might have applications in quantum computing [35,36].

**Methods:**
**Etching process on $h$-BN**
First, the $h$-BN flakes were mechanically exfoliated on quartz substrates. The substrates were then annealed at 650 °C in a $O_2$ flow for 60 min and a subsequent Ar:$H_2$ flow for another 60 min to clean the surface of $h$-BN. Next, a $NiCl_2$ solution in 0.1 mg ml$^{-1}$ or an $H_2PtCl_6$ solution in 10 ml L$^{-1}$ was spun at 4,000 r.p.m. for 100 seconds onto the substrate. The substrates were then baked for 10 min on a hot plate. Then samples coated with $NiCl_2$ were submitted to a two-step process: annealing at 1,000°C for 30 min under an Ar:$H_2$ flow (10:2 sccm) and etching at 1,200°C for 30-60 min under an Ar:$H_2$ flow (850:150 sccm). The etching pressure was found to be ~150 Pa. The samples with $H_2PtCl_6$ were annealed at 1,150 °C for 30 min, and then kept at 1,300°C for 10-30 min in an Ar:$H_2$ flow (30:10 sccm) under a pressure of ~7 Pa. After etching, the system was cooled in the protection of argon flow [See also Supplementary **Fig.** S1].

**GNR growth**
After nano-trenches were fabricated in *h*-BN, the quartz substrates with *h*-BN flakes were subjected to ultrasonic bathing in an HCl solution, DI water and acetone, in sequence. Subsequently, the quartz substrates were loaded to a furnace and heated to 1,280 °C in argon flow, and after that, a $C_2H_2$ flow and a mixture of silane/argon (5% mole ratio of silane) were introduced into the chamber for GNR growth. The ratio of $C_2H_2$ to silane was optimized for AGNR and ZGNR growth, respectively. The pressure was kept at ~2 Pa during ZGNR growth and ~1 Pa during AGNR growth, and the typical time for growth is 2 min. After growth, the substrates cooled down in the protection of argon flow [also see Supplementary **Fig.** S2]. The *h*-BN flakes with ultra-narrow GNRs were transferred onto silicon substrate capped with 300 nm of $SiO_2$ for further transport measurement.

**Atomic force microscopy (AFM)**
Both *h*-BN nano-trenches and GNR samples were characterized by an AFM (Dimension Icon, Bruker) in contact mode, while the atomic-resolution images were obtained by another AFM (Multimode IV, Veeco) under ambient conditions. The AFM images were all recorded in contact mode using SNL-10 AFM tips from Bruker that possess a nominal tip radius of less than 10 nm. For atomic-resolution scanning, the force constant *k* of the cantilever tips was in the range of 0.05-0.5 N $m^{-1}$, and the scan rate was set to a value in the range of 10–60 Hz to reduce the noise from thermal drift. To obtain a high accuracy, several hours of pre-scanning were carried out to warm up the scanner and ensure high imaging stability, and scanners with a travel range less than 10 μm along the X and Y directions were used.

**Raman spectroscopy characterization**
Raman spectra were obtained with a WITec micro-Raman instrument (mode Alpha 300R) possessing excitation laser lines of 532 nm. An objective lens with x100 magnification and a 0.95 numerical aperture was used, producing a ~500 nm in diameter laser spot. The laser power was kept less than 1 mW on the sample surface to avoid laser-induced heating.

**Computational details of thermodynamic simulation for *h*-BN edges and GNR growth**
The calculation was performed within the framework of DFT as implemented in the Vienna Ab initio Simulation Package (VASP). The electronic exchange and correlation were included through the generalized gradient approximation (GGA) in the Perdew-Burke-Ernzerhof (PBE) form. The interaction between valence electrons and ion cores was described by the projected augmented wave (PAW) method and the energy cutoff for the plane wave functions was 400 eV. All structures were optimized by a conjugate gradient method until the energy was converged to $1.0 \times 10^{-5}$ eV/atom. The climbing image nudged-elastic band (CI-NEB) method was utilized to search for the transition states involved in the graphene growth, and the minimum energy path was optimized using a force-based conjugate-gradient method until the force was converged to 0.02 eV/Å. The vacuum layer inside the super-cell was kept to be larger than 15 Å to avoid the interaction between adjacent unit cells. For graphene growth at AC and ZZ edges, supercells of 13 Å× 28 Å× 15 Å and 12.565 Å× 28 Å× 15 Å were built and a k-point mesh of 2×1×1 was used.

**Sample preparation for TEM and STEM investigation**
To fabricate very thin *h*-BN flakes for further TEM investigation, we cleaved *h*-BN flakes grown with GNRs onto top of an oxidized Si wafer. Thin *h*-BN flakes (less than 2.2 nm) were selected with the help of an optical microscope and an AFM. The thin flakes were then transferred onto a gold Quantifoil (R) TEM grid, where portions of the hetero-structure are freely suspended on holes measuring approximately ~1 μm in diameter.

### TEM and STEM characterization

Initial characterization was performed in a Philips CM200 microscope operated at 80 kV. Selected samples were characterized at high resolution using an aberration-corrected Nion UltraSTEM100 operated at 60 kV. These samples were annealed via radiative or laser-induced heating before STEM measurements to reduce the contaminant on the surface.

### Device fabrication and electrical measurements

GNR transistors were fabricated by a standard e-beam lithographic technique with 20 nm Pd as source/drain contacts. The thickness of *h*-BN flakes transferred on Si substrate with 300 nm SiO$_2$ capping layer is normally in a range from 15 to 35 nm. The devices were then annealed in a hydrogen flow at 200 °C for 3 hours to improve the contact quality. The electrical transport was measured in PPMS (Quantum Design, Inc.) at pressure ~1 × 10$^{-6}$ torr via Keithley 4200.

**Acknowledgments：**
H.W. and X.X. thank J.H. Edgar (Kansas State University, USA) for supplying the partial *h*-BN crystals. H. S. Wang, L. Chen and H. Wang thank M. Liu, X. Qiu and J. Pan from NCNT of China, F. Liou, H. Tsai, M. Crommie from UCB, USA, J. Xue and P. Yu from ShanghaiTech. University and S. Wang from SJTU for nc-AFM measurement. H. S. Wang, L. X. Chen and H. Wang thank B. Sun and S. Li from Hunan University for the fusion of the STEM image and the EELS mapping images. **Funding:** The work was partially supported by the National Key R&D program (Grant No. 2017YFF0206106), the Strategic Priority Research Program of Chinese Academy of Sciences (Grant No. XDB30000000), the National Science Foundation of China (Grant No. 51772317, 51302096, 61774040), the Science and Technology Commission of Shanghai Municipality (Grant No. 16ZR1442700, 16ZR1402500 18511110700), Shanghai Rising-Star Program (A type) (Grant No.18QA1404800), the Hubei Provincial Natural Science Foundation of China (Grant No. ZRMS2017000370), China Postdoctoral Science Foundation (Grant No. 2017M621563, 2018T110415), and the Fundamental Research Funds of Wuhan City (No. 2016060101010075). C.L. acknowledges support from the European Union's Horizon 2020 research and innovation programme under the Marie Skłodowska-Curie grants No. 656378 – Interfacial Reactions. T.J.P. acknowledges funding from European Union's Horizon 2020 Research and Innovation Programme under the Marie Sklodowska-Curie grant agreement no. ~655760 -- DIGIPHASE. K.W. and T.T. acknowledge support from the Elemental Strategy Initiative conducted by the MEXT, Japan and the CREST (JPMJCR15F3), JST. C.X.C. acknowledges financial support from the National Young 1000 Talent Plan of China and the National Key R&D Program of China (No. 2018YFA0703700). L.H. acknowledges financial support from the program of China Scholarships Council (No. 201706160037).

**Author Contributions：**
H.W. and X.X. directed the research work. H.W. conceived and designed the research. L.H., H.S.W and C.C. performed the etching processes on *h*-BN. T.W. suggested the use of platinum as cutting particle. L.C. and C.J. performed the growth experiments for the GNRs. L.C., C.C., C.J. and H.S.W. performed AFM measurements. H.S.W. fabricated the electronic devices and performed the transport measurements. C.X.C. performed the Raman measurements. K.E., C.L., T.J.P., G.A. and J.C.M. carried out STEM measurements. K.W. and T.T. fabricated the *h*-BN crystals. W.W. and Q.Y. performed the thermodynamic simulation for *h*-BN edges and GNR growth. H.W., H.S.W., L.C., J.C.M., K.E., L.H. and C.X.C. analyzed the experimental data and designed the figures. H.W., H.S.W. and J.C.M. co-wrote the manuscript, and all authors contributed to critical discussions of the manuscript.

# Supplementary Materials

## Supplementary Figures

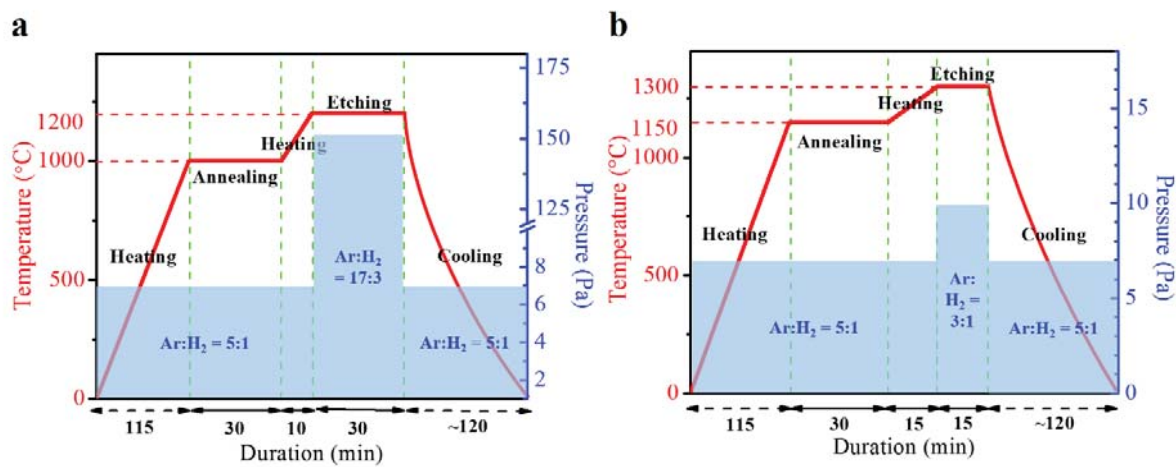

**Fig. S1. Schematic diagrams of typical processes for catalytic etching on *h*-BN. (a)** A typical etching process for ZZ oriented trenches. The $NiCl_2$-coated samples were submitted to a two-step process: annealing at 1,000 °C for 30 min and etching at 1,200 °C for 30 min under an $Ar:H_2$ flow (850:150 sccm), and the etching pressure was kept at ~150 Pa. **(b)** A typical etching process for AC oriented trenches. The $H_2PtCl_6$-coated samples were annealed at 1,150 °C for 30 min and etched at 1,300 °C for 15 min under an $Ar:H_2$ flow (30:10 sccm) with the etching pressure ~10 Pa. The pressure during heating, annealing and cooling process for both system was kept at ~7 Pa with an $Ar:H_2$ (10:2 sccm) flow.

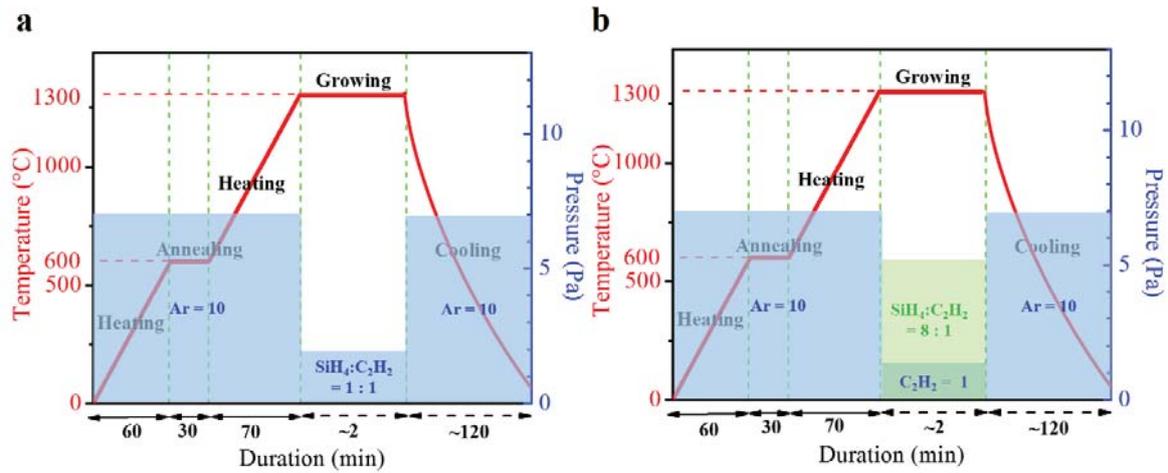

**Fig. S2. Schematic diagrams of edge-specific GNR growth on *h*-BN. (a)** A typical process for ZGNR growth. Before GNR growth, the samples were first heated to 600 °C to remove contaminants. The ratio of silane to $C_2H_2$ was ~1:1, the pressure was maintained at ~2 Pa during the growth. **(b)** A typical process for AGNR growth. After heating, $C_2H_2$ gas (or a mixture of $SiH_4$ to $C_2H_2$ was ~8:1) was fed in for AGNR growth and the pressure was kept at ~1.6 (~5.2) Pa. During heating, annealing and cooling, the pressure was kept at ~7 Pa with an Ar flow of 10 sccm for all samples.

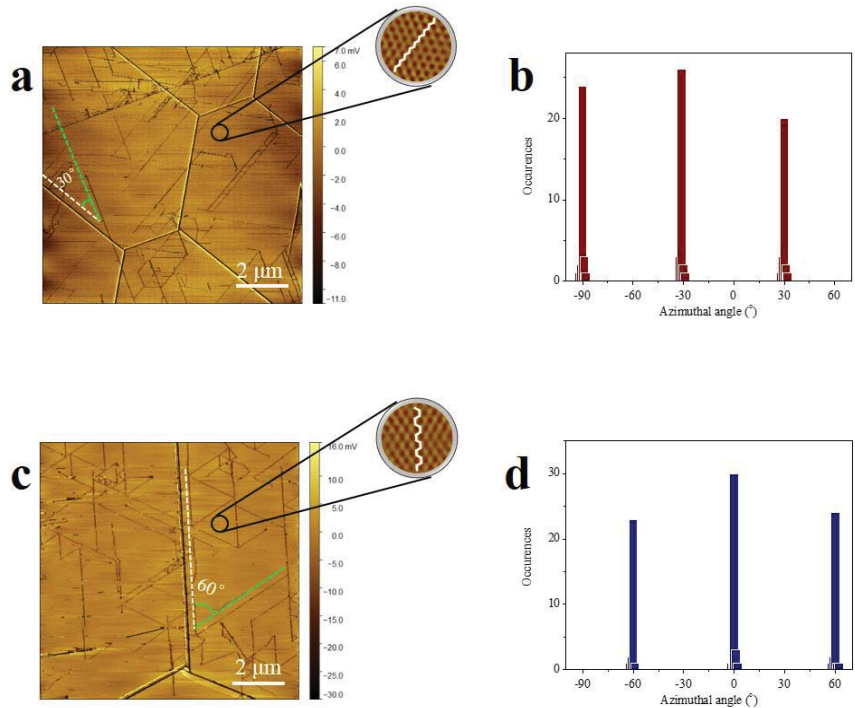

**Fig. S3. AFM investigation on *h*-BN nano-trenches obtained via nickel and platinum nano-particle etching.** **(a)** AFM friction image of zigzag oriented trenches produced by nickel particles. The trenches exhibit about -30°, 30° or 90° with respect to *h*-BN wrinkles which are armchair oriented. The green dash line represents a trench orientation while the white dash line denotes a wrinkle orientation. Inset is a zoom-in view of the selected region in-lattice-resolution, confirming that the trenches are along the zigzag direction. The lattice-resolution AFM friction image is Fourier filtered for clarity. **(b)** Occurrences of all trenches obtained via nickel assisted etching versus their azimuthal angle. Histogram shows that three specific angles (-30°, 30° and 90°) are preferred. It indicates that zigzag direction dominates in the trenches etched by Ni nanoparticles. **(c)** AFM friction image of armchair oriented trenches on *h*-BN cut by platinum particles. The trenches exhibit about -60°, 0° or 60° with respect to the *h*-BN wrinkles. The green dash line represents a trench orientation while the white dash line denotes a wrinkle orientation. Inset is a zoom-in view of the selected region shown in (c), showing Fourier filtered lattice-resolution AFM friction image, confirming that the trenches are along the armchair direction. **(d)** Occurrences of all trenches obtained via platinum assisted etching versus their azimuthal angle. Histogram shows three specific angles (-60°, 0° and 60°) are preferred by the Pt assisted etching. It means that armchair orientation dominates in all trenches etched by Pt nanoparticles.

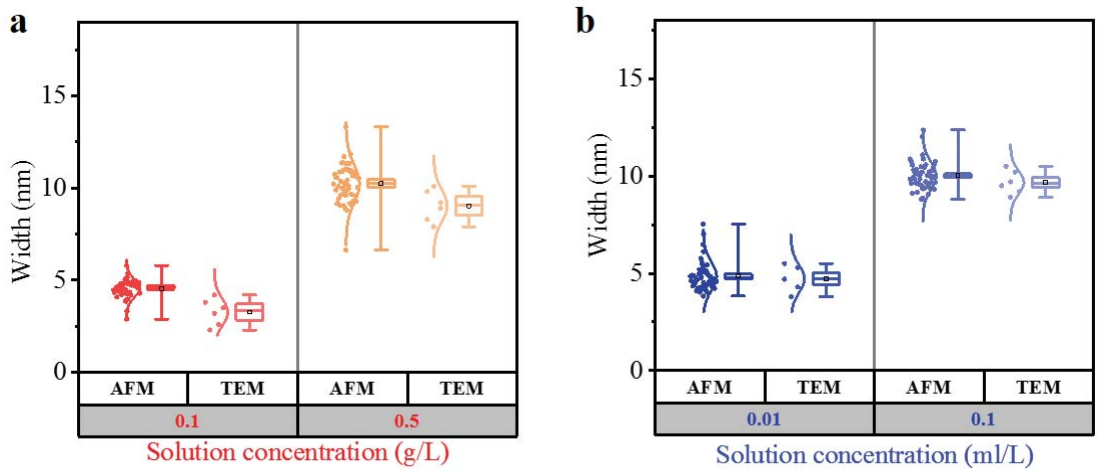

**Fig. S4. Boxplots of trench width distributions under optimized etching conditions. (a)** Width distribution of the zigzag-trenches obtained at 1200 °C. **(b)** Width distribution of armchair trenches obtained at 1300 °C.

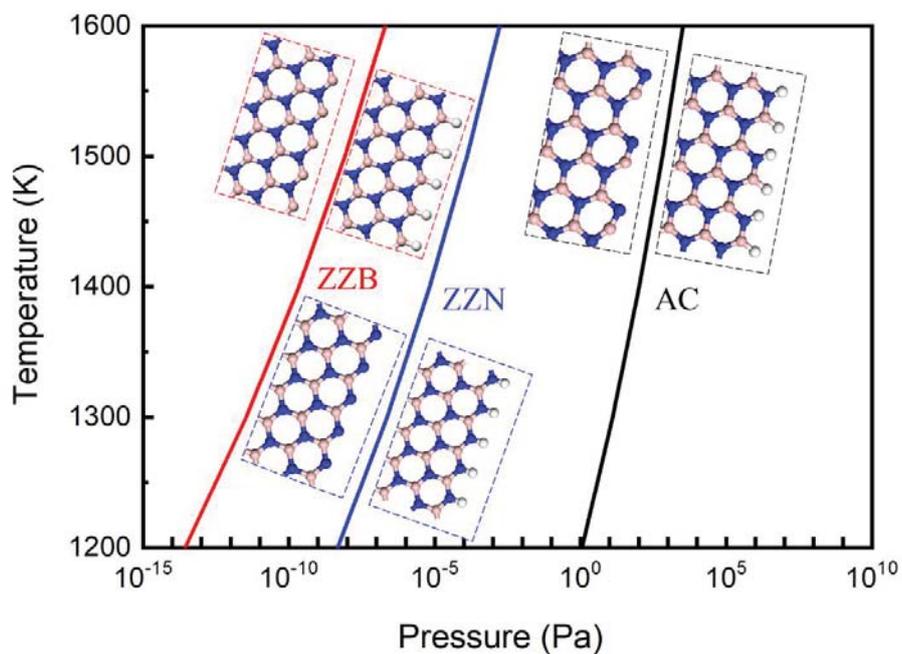

**Fig. S5. Thermodynamic diagrams of *h*-BN edges with configurations of zigzag boron (ZZB), zigzag nitrogen (ZZN) and armchair (AC).** The edge structure of *h*-BN is highly dependent on the temperature and hydrogen pressure. Edges at the left side of the phase transition line are non-passivated while edges at the right side of line is hydrogen passivated. ZZB edge is easily to be passivated even at very low hydrogen pressure. AC edge is very stable and only very high hydrogen pressure can passivate it.

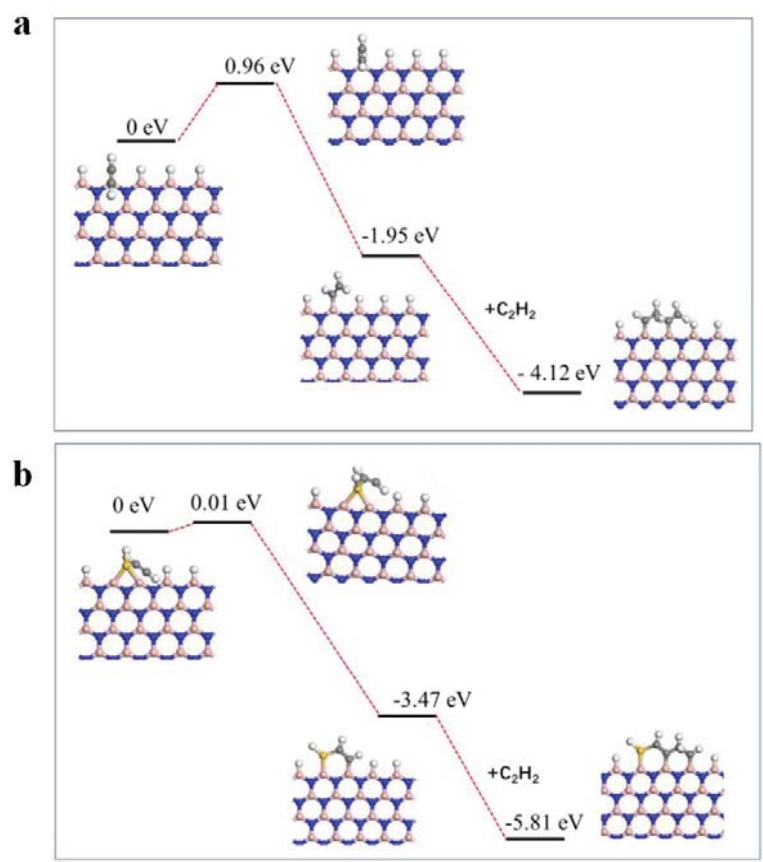

**Fig. S6. Energy profiles for the growth of GNR at ZZB edge without and with the catalyst of silane. (a)** GNR growth on H-passivated ZZB edge of *h*-BN. **(b)** GNR growth at the H-passivated ZZB edge of *h*-BN by using silane as a catalyst. The using of silane catalyst greatly decreases the activation barrier of $C_2H_2$ incorporation into the edge.

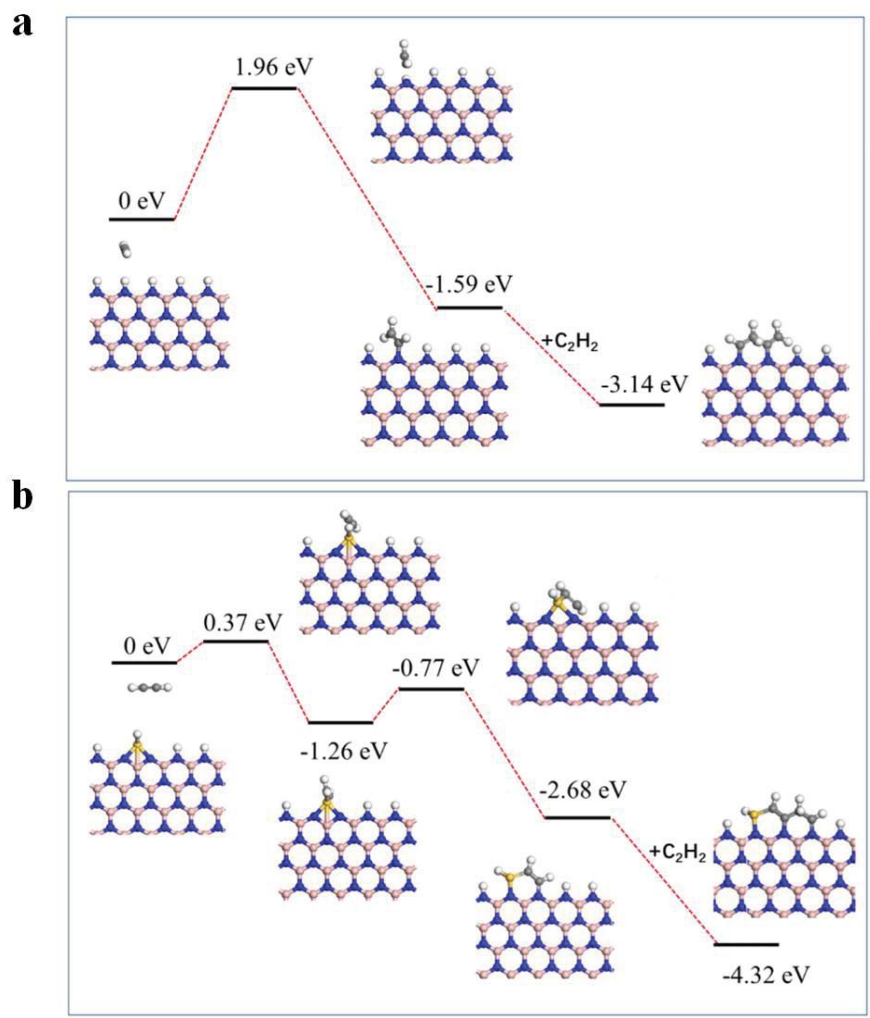

**Fig. S7. Energy profiles for the growth of GNR at ZZN edge without and with the catalyst of silane. (a)** GNR growth on H-passivated ZZN edge of *h*-BN. **(b)** GNR growth at the H-passivated ZZ edge of *h*-BN by using silane as a catalyst. The using of silane catalyst greatly decreases the activation barrier of $C_2H_2$ incorporation into the edge.

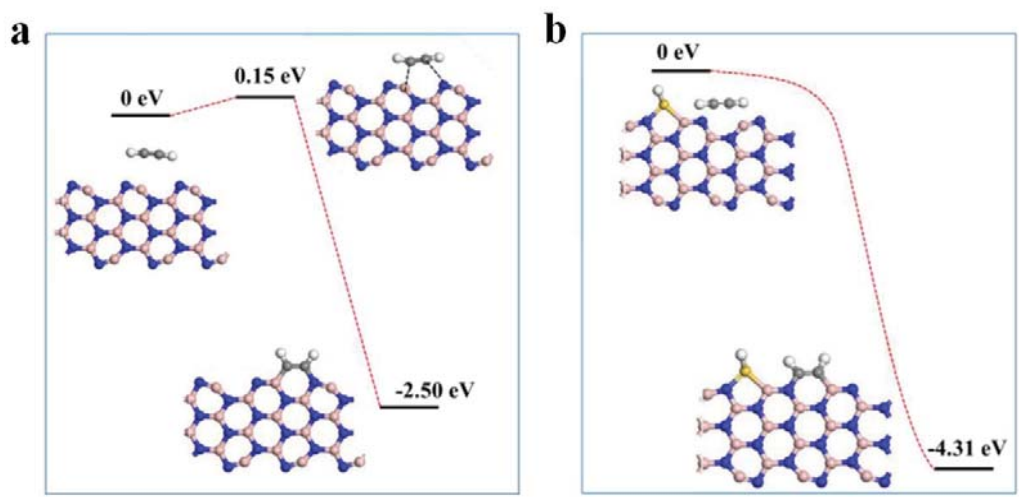

**Fig. S8. Energy profiles for the growth of GNR at AC edge without and with the catalyst of silane. (a)** GNR growth at the AC edge of *h*-BN. **(b)** GNR growth at the AC edge of *h*-BN by using silane as a catalyst. Both edges have very low activation barrier to the addition of $C_2H_2$.

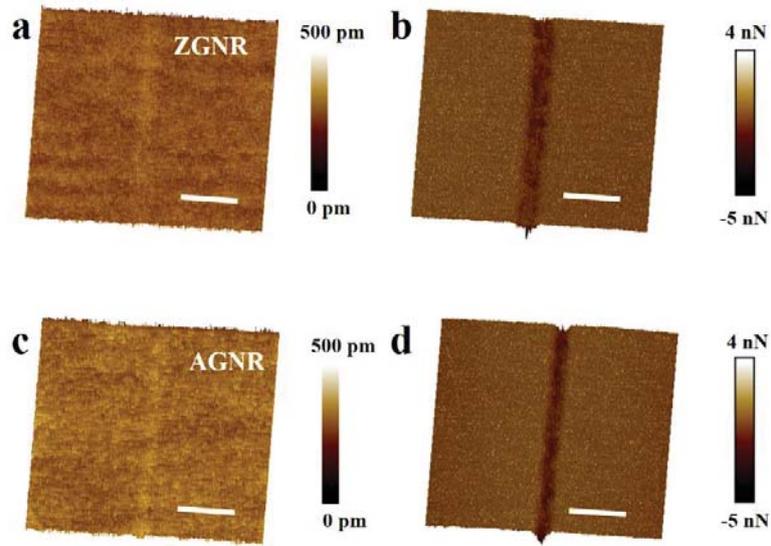

**Fig. S9. AFM images of ultra-narrow GNRs embedded within *h*-BN substrate.** **(a)** 3D AFM height image and **(b)** 3D friction image of a sub-5 nm-wide ZGNR on an *h*-BN substrate. **(c)** 3D AFM height image and **(d)** 3D friction image of a sub-5 nm-wide AGNR in an *h*-BN substrate. The scale bars are 20 nm.

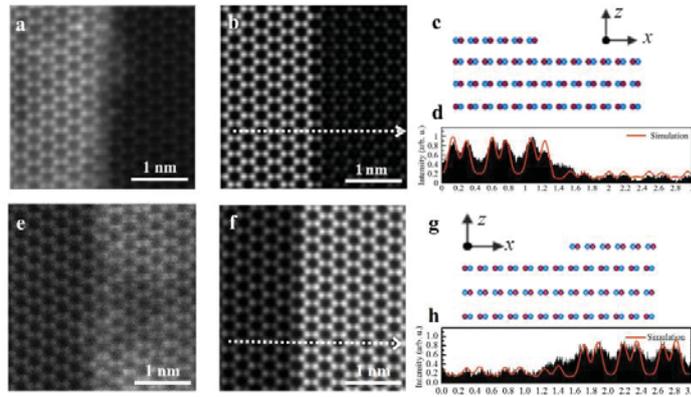

**Fig. S10. Atomic resolution analysis on the zigzag edge of *h*-BN trench. (a,e)** Wiener filtered HAADF-STEM images of *h*-BN edge. **(b,f)** Simulated HAADF images of the models corresponding the experimental images. **(c,g)** Top views of the rigid models showing the etched trench of *h*-BN crystal consisting of 4 layers. In the model, blue and red correspond to Nitrogen and Boron, respectively. **(d,h)** Intensity profiles recorded over the dashed line on panel b and f, respectively.

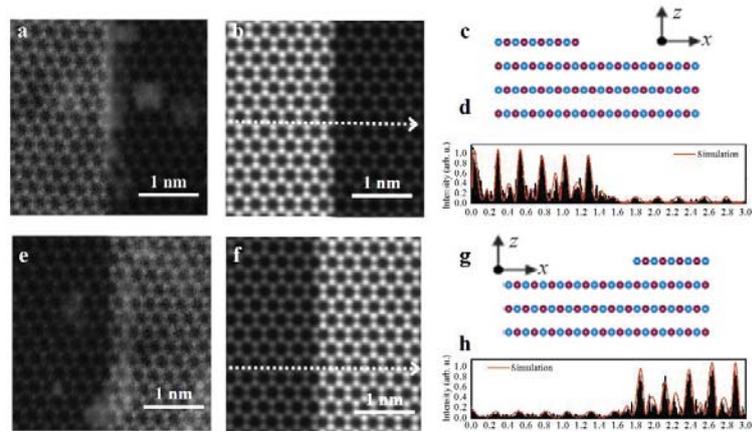

**Fig. S11. Atomic resolution analysis on the armchair edge of *h*-BN trench. (a,e)** Wiener filtered HAADF-STEM images of *h*-BN edge. **(b,f)** Simulated HAADF images of the models corresponding the experimental images. **(c,g)** Top views of the rigid models showing the etched trench of *h*-BN crystal consisting of 4 layers. In the model, blue and red correspond to Nitrogen and Boron, respectively. **(d,h)** Intensity profiles recorded over the dashed line on panel b and f, respectively.

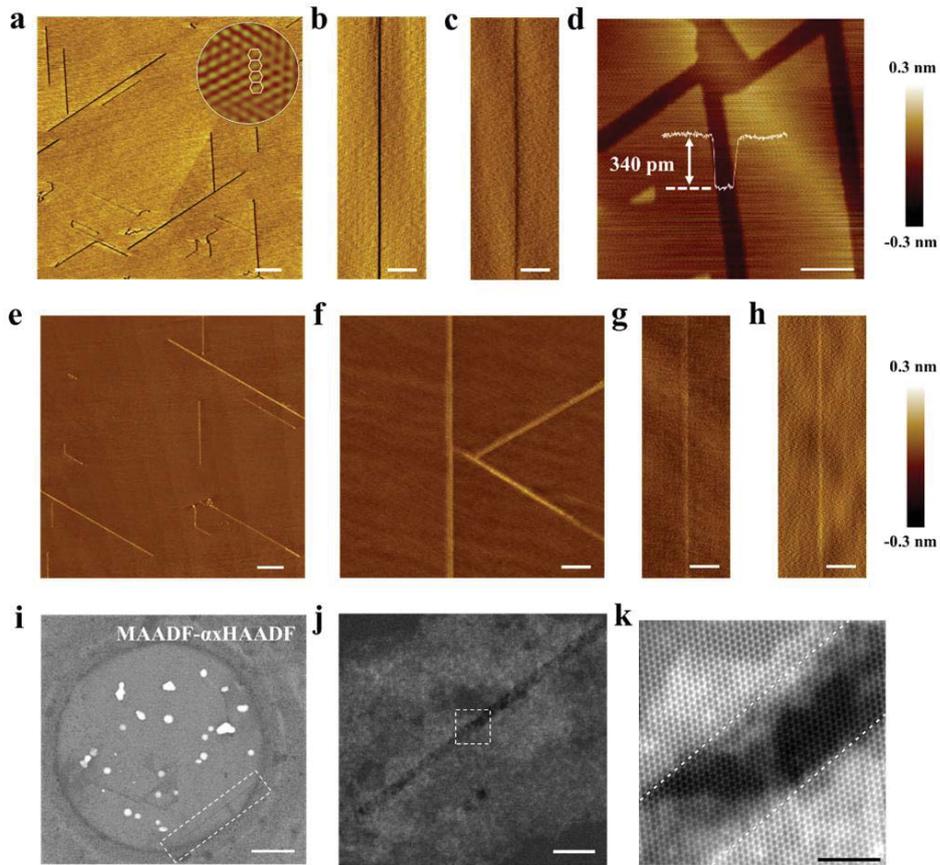

**Fig. S12. High resolution analysis of ZZ-oriented nano-trenches and ZGNRs embedded in *h*-BN lattices.**
**(a-d)** AFM height images of mono-layered ZZ-oriented nano-trenches in *h*-BN surface by Ni particle-catalyzed cutting. The scale bars are 500, 100, 100 and 200 nm, respectively. The circular inset in **(a)** shows an atomic-resolution friction image of the *h*-BN. All the trenches are found along ZZ direction. **(b)** and **(c)** show nano-trenches narrower than 5 nm. **(d)** For a ~78 nm-wide trench, the depth profile shows that etching occurs only at the top layer of *h*-BN substrate. **(e-h)** AFM height images of GNRs embedded in the ZZ-oriented nano-trenches. The width of ZGNRs is less than 10 nm. The scale bars are 500, 200, 100 and 100 nm, respectively. **i** A STEM MAADF-α×HAADF image for a ZGNR sample. The scale bar is 200 nm. **(j)** A zoomed-in MAADF image of a region shown in the middle of white dashed frame in **(i)**. The scale bar is 10 nm. **(k)** A Wiener-filtered MAADF image of the region shown in the dashed frame in **(j)**. The STEM investigation indicates that the boundaries between GNR and *h*-BN can be distinguished with a scale bar of 2 nm. The measured in-plane width of the GNR is ~3.2 nm.

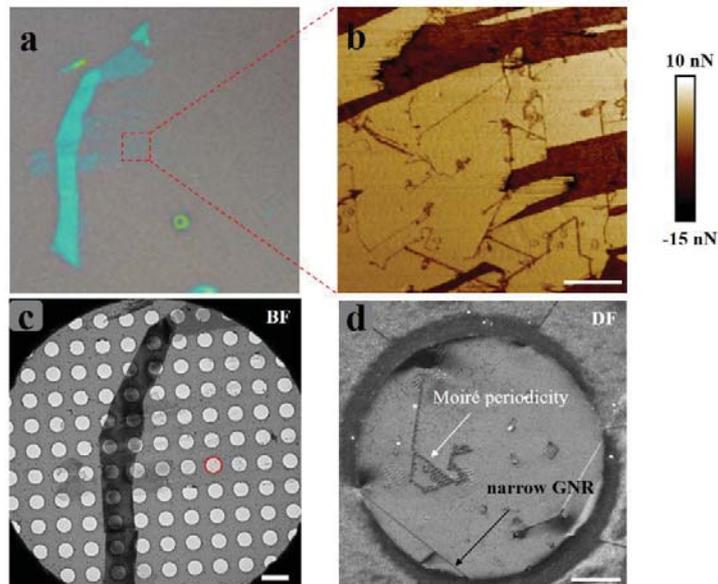

**Fig. S13. TEM sample preparation and characterization of GNRs embedded in *h*-BN. (a)** Optical image of *h*-BN flakes exfoliated on a Si substrate with 300 nm SiO$_2$. **(b)** The AFM friction image taken in the area of the red dashed frame shown in (a), scale bar 600 nm. The dark stripes are GNRs. The thickness of the thin *h*-BN flake is in the range of 1.3-2.2 nm, which corresponds to 3-5 layers of *h*-BN. **(c)** TEM bright field (BF) image of the sample shown in (a) after being transferred onto a TEM grid. The scale bar is 3 μm. **(d)** TEM dark field (DF) image of the region inside the red circle in (c). The black arrow points to a narrow GNR. The white arrow points to a ~14 nm Moiré periodicity, which corresponds to the lattice mismatch between graphene and *h*-BN. The scale bar is 200 nm.

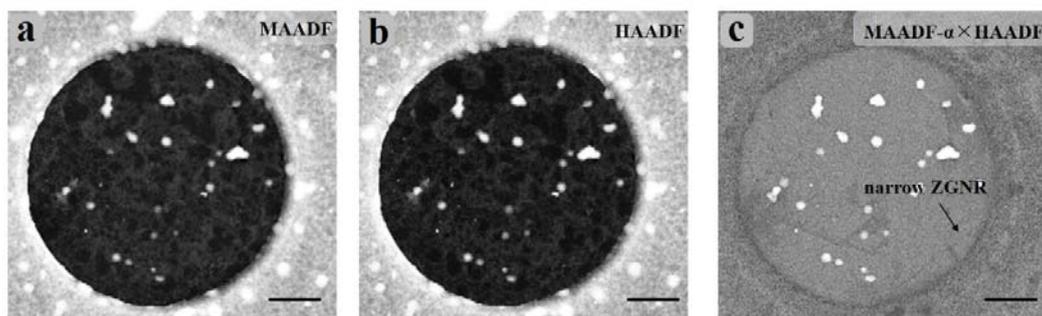

**Fig. S14. The STEM images of GNR. (a)** Medium angle annular dark field (MAADF). **(b)** High angle annular dark field (HAADF). **(c)** MAADF-α×HAADF images, α is a variable value which is adjusted so as to minimize the contrast of adsorbed contamination. The arrow points to the narrow ZGNR corresponding to the same ZGNR shown in **Fig.** S12i-k and **Fig.** S13d. The scale bars are 200 nm.

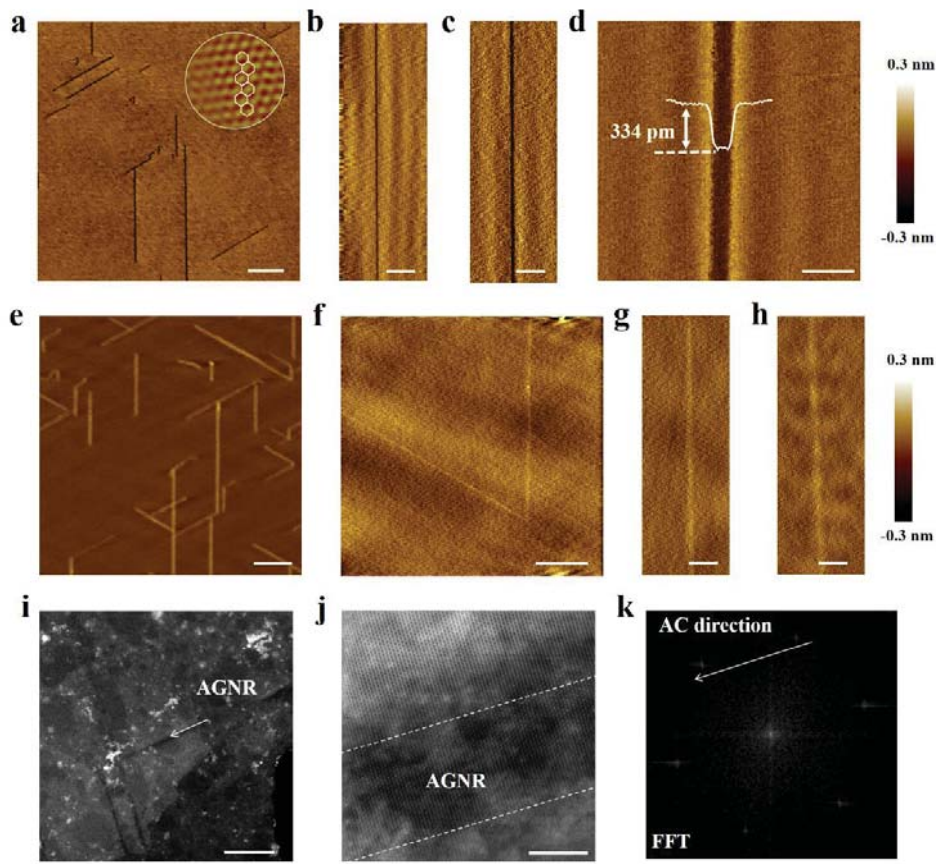

**Fig. S15. High-resolution characterization of AC-oriented nano-trenches and AGNRs embedded in *h*-BN trenches. (a-d)** AFM height images of mono-layered AC-oriented trenches obtained via Pt particle-assisted etching. The scale bars are 500, 100, 100 and 50 nm, respectively. The inset circular in **(a)** shows atomic-resolution friction image of *h*-BN. The white hexagons in the inset help with the identification of the atomic structure of GNR. All the trenches are found along AC directions. The width of the trench is less than 5 nm. **(d)** shows an AC-oriented trench in width of ~23 nm, the profile of the depth indicates that etching occurs only at the top layer of *h*-BN surface. **(e-h)** AFM height images of AGNRs embedded in the AC nano-trenches. The scale bars are 500, 200, 100 and 100 nm, respectively. The width of AGNRs is less than 10 nm. **(i)** STEM-MAADF images of an AGNR sample. **(j)** The magnified image of the AGNR area pointed to by the arrow in **(i)**, and the two dashed lines show the boundary between *h*-BN and AGNR. The scale bar in **(i)** is 100 nm and in **(j)** is 3 nm. The width of the GNR is ~5.3 nm. **(k)** The fast Fourier transform (FFT) image for **(j)**, indicating that the GNR is along the AC lattice direction.

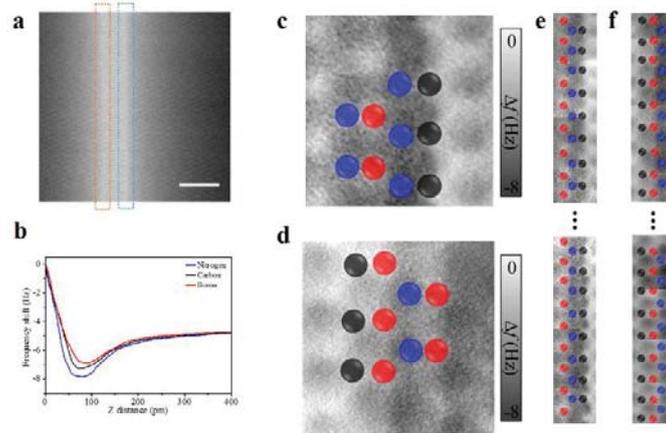

**Fig. S16. High-resolution AFM investigation on ZGNR. (a)** A typical nc-AFM image of a ZGNR embedded in *h*-BN in constant-height mode. Scale bar: 4 nm. **(b)** Force spectroscopies taken at different elements in the embedded ZGNR. **(c)** A zoom-in view of one boundary of ZGNR-*h*-BN, here boron in red, carbon in black and nitride in blue. **(d)** A close view of the other edge of ZGNR-*h*-BN. **(e-f)** A series of nc-AFM images which were continuously measured along the opposite boundaries in the orange (nitride terminated) and blue (boron terminated) box indicated in (a), respectively. The nc-AFM measurement parameters: $V_{tip}$ = 0 mV while the amplitude was kept at ~60 pm.

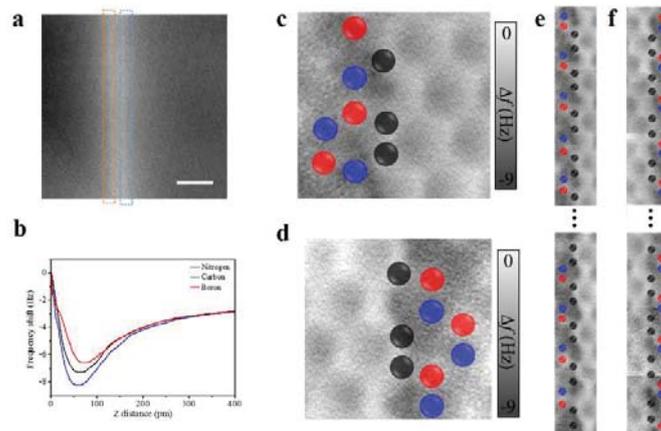

**Fig. S17. High-resolution AFM results of AGNR.** (a) A typical constant-height nc-AFM image of an AGNR embedded in *h*-BN. Scale bar: 10 nm. (b) Force spectroscopies taken at different elements in the embedded AGNR. (c-d) Zoom-in views of the boundaries of AGNR-*h*-BN with boron in red, carbon in black and nitride in blue. (e-f) A series of nc-AFM images corresponding to orange and blue box in (a). The parameters for nc-AFM measurement: $V_{tip}$ = 0 mV while the amplitude was kept at ~60 pm.

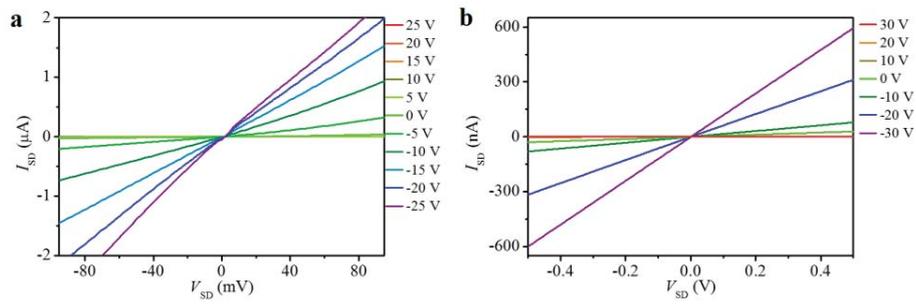

**Fig. S18.** $I_{SD}$-$V_{SD}$ characteristics recorded under different $V_{gate}$ when **(a)** $T = 100$ K for the device shown in **Fig.** S21 and **(b)** $T = 150$ K for the device shown in **Fig.** S19.

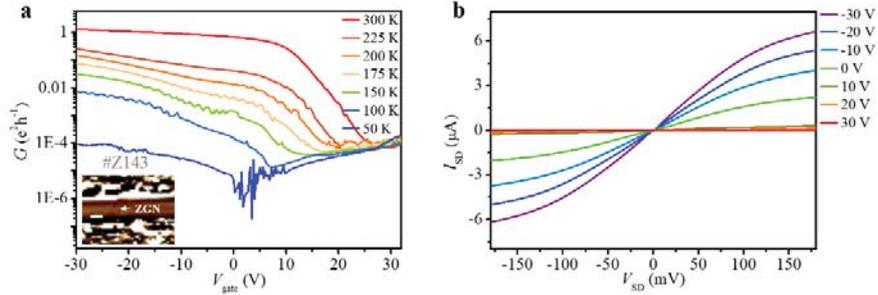

**Fig. S19. Electronic transport through a sub-5 nm ZGNR device on *h*-BN. (a)** Gate voltage dependence of the conductance (*G*) of a ZGNR with an estimated width of about 5 nm. The conductance can be switched off even at room temperature. The inset shows an AFM friction image of the nanoribbon channel corresponding to the transfer curves. The channel length is about 247 nm. The scale bar in the inset is 200 nm. Its field effect mobility is about 1,639 cm$^2$V$^{-1}$s$^{-1}$ at 300 K. The GNR field effect transistor (FET) with a channel length of about 247 nm exhibits a rather high $G = 1.3$ $e^2h^{-1}$ at room temperature when $V_{gate}$= -30 V. Charge carrier was scattered for (4 $e^2h^{-1}$)/(1.3 $e^2h^{-1}$)≈3 times. The scattering mean free path (MFP) is estimated to be about 247 nm/(3+1) = 62 nm even without excluding the contribution from the contact. The extracted band gap by STB model is 436.2±28.1 meV. **(b)** $I_{SD}$-$V_{SD}$ curves for the same device in (a) at different gate bias voltages at room temperature.

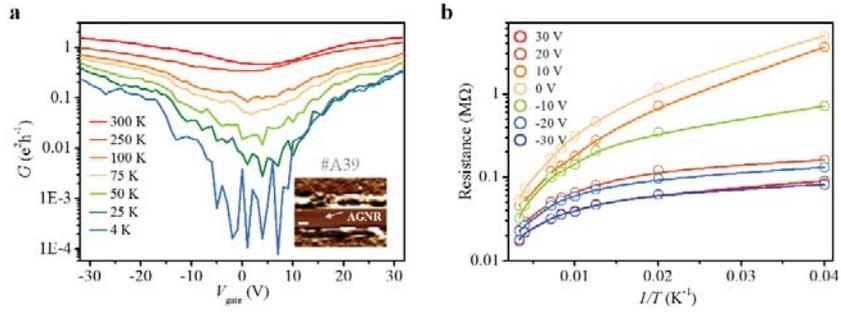

**Fig. S20. Electronic transport through a sub-5 nm AGNR device on *h*-BN. (a)** Conductance (*G*) of an armchair graphene nanoribbon (AGNR) with a width of ~5 nm as a function of back gate voltage ($V_{gate}$). The inset shows an AFM friction image of the AGNR channel corresponding to the transfer curves. The scale bar is 200 nm. The channel length is about 278 nm. Its field effect mobility is about 1,700 cm$^2$V$^{-1}$s$^{-1}$ at 300 K. **(b)** Resistance (*R*) under different $V_{gate}$ versus temperature (*T*) from 25 to 300 K for the 5 nm-wide AGNR, the solid curves are fits based on a simple two-band model, the extracted band gap is about 183.2±18.7 meV.

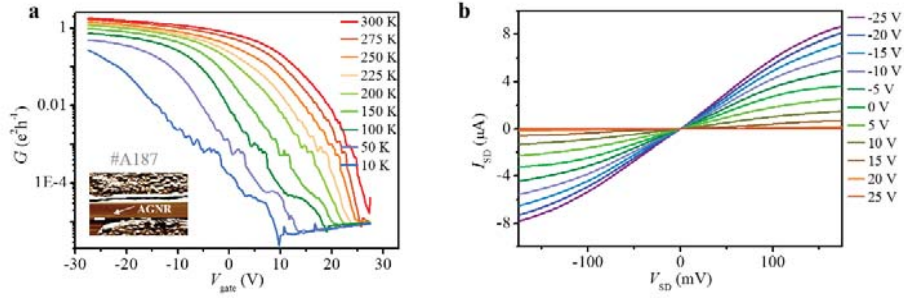

**Fig. S21. Electronic transport through a sub-5 nm AGNR device on *h*-BN. (a)** Conductance (*G*) of an armchair graphene nanoribbon (AGNR) with an estimated width of about 4.8 nm. The conductance can be completely switched off even at room temperature. The inset shows the AFM friction image of the nanoribbon channel corresponding to the transfer curves. The scale bar in the inset is 200 nm. Its field effect mobility is about 1,428 cm$^2$V$^{-1}$s$^{-1}$ at 300 K. The GNR field effect transistor (FET) in a channel length of about 236 nm exhibits rather high $G = 1.8\ e^2h^{-1}$ at room temperature when $V_{gate}$ = -25 V. Charge carrier was scattered for ($4e^2h^{-1}$)/($1.8\ e^2h^{-1}$)≈2 times. The scattering mean free path (MFP) is estimated to be about 236 nm/(2+1) = 79 nm even without excluding the contribution from the contact. The extracted band gap by STB model is 530.9±34.7 meV. **(b)** $I_{SD}$-$V_{SD}$ curves for the device in (a) at different gate bias voltages at room temperature.

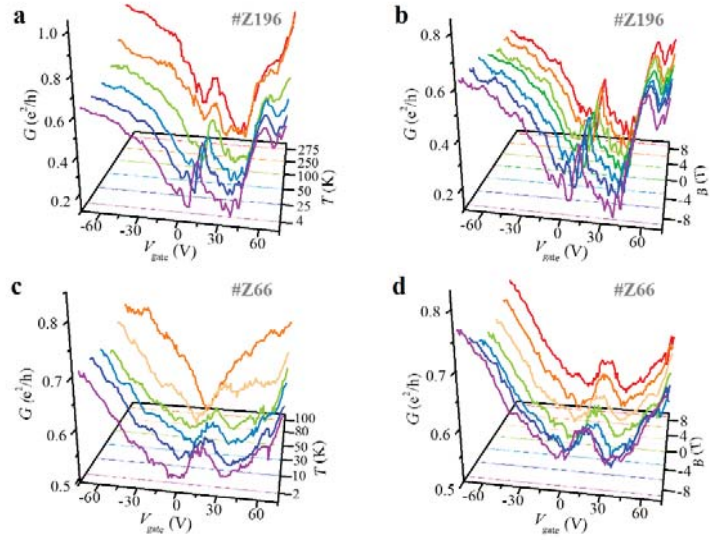

**Fig. S22. The influence of magnetic field and temperature on transport properties of ZGNRs.** Transfer curves of **(a)** sample #Z196 and **(c)** sample #Z66 at different temperatures. **(b)** Transfer curves for sample #Z196 subjected to different magnetic field ($B$) at the temperature of 4 K. **(d)** The typical transfer curves for sample #Z66 at several $B$ at 2 K. All the measurements were under $V_{SD}$ = 20 mV.

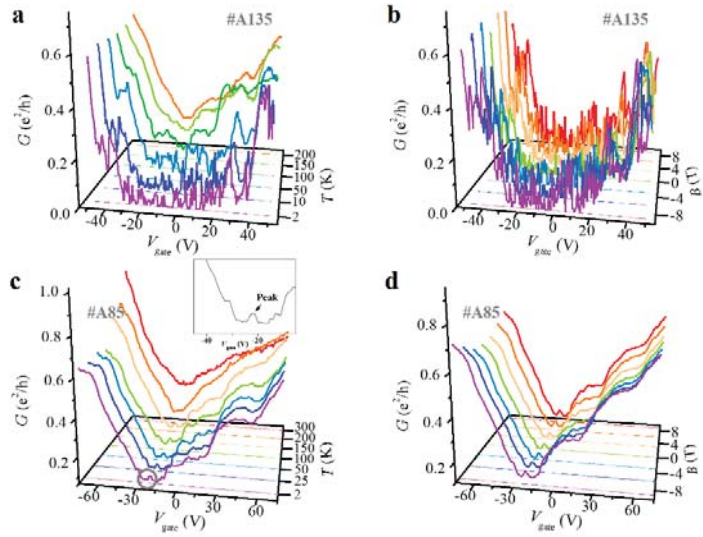

**Fig. S23**. **The influence of magnetic field and temperature on transport properties of AGNRs.** The typical transfer curves for **(a)** sample #A135 and **(c)** sample #A85 at several temperatures. The inset shows enlarged area in the grey circle in (c). The typical transfer curves for **(b)** sample #A135 and **(d)** sample #A85 at several $B$ at 2 K. All the measurements were under $V_{SD}$ = 10 mV.

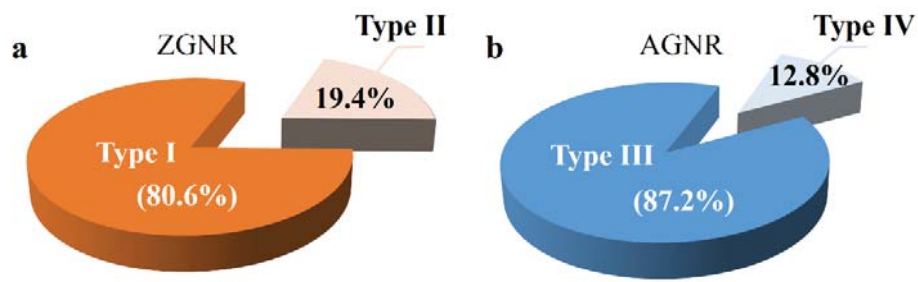

**Fig. S24. The investigation on conductance peak in GNRs. (a)** Pie chart of distribution of different ZGNRs. Type I corresponds to ZGNRs whose peaks always exist even at high temperature (a typical example: #Z196), and Type II represents ZGNRs whose peaks disappear at high temperature (a typical example: #Z66). **(b)** Pie chart of distribution of different AGNRs. Type III is AGNR whose transfer curves are in absence of nontrivial conductance peaks at any temperature (a typical example: #A135) and Type IV means AGNR whose transfer curve is of a tiny peak at low temperature (a typical example: #A85).

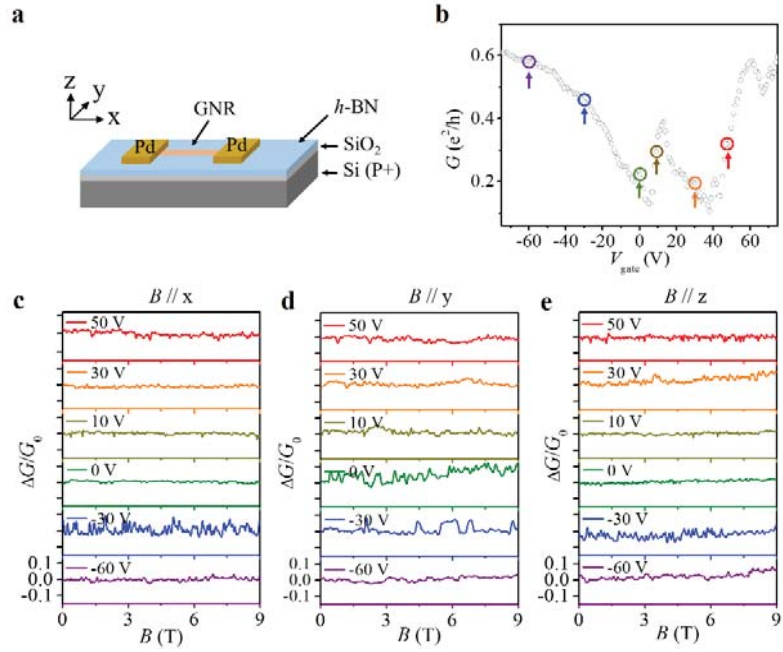

**Fig. S25. Magneto-electronic properties of a typical Type I ZGNR FET (sample #Z196) under different $V_{gate}$.** **(a)** Schematic of a ZGNR device. **(b)** Transfer characteristics of the $8.9 \pm 0.5$ nm ZGNR FET. **(c-e)** Normalized MC measured at selected $V_{gate}$ indicated by the colored dots marked on the transfer curve shown in (a), the magnetic field was applied (c) in parallel with the direction of the current, (d) in the plane of the substrate, but perpendicular to the current, and (e) perpendicular to the substrate plane. All the measurements were carried out at 4 K with $V_{SD} = 20$ mV.

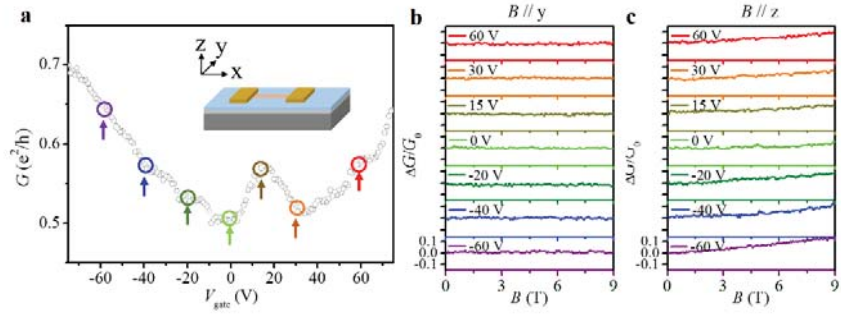

**Fig. S26. Magneto-electronic properties of a typical Type II ZGNR FET (sample #Z66). (a)** Transfer characteristics for the 9.2 ± 0.5 nm ZGNR with Pd contacts. Inset: Schematic of the ZGNR device. Normalized MC measured at different $V_{gate}$ indicated by the colored dots marked on the transfer curve in (a) when the magnetic field was applied **(b)** in the plane of the substrate but perpendicular to the current, or **(c)** perpendicular to the substrate. All the measurements were carried out when $V_{SD}$ = 20 mV and $T$ = 2 K.

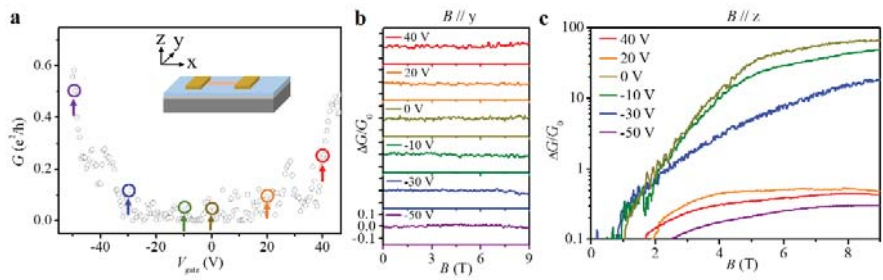

**Fig. S27. Magneto-electronic properties of Type III AGNR FET (sample #A135). (a)** Transfer characteristics for the 9.5 ± 0.5 nm AGNR FET, colored circles (and the arrows) denote certain $V_{gate}$ where magneto-conductance (MC) was measured; Normalized magneto-conductance measured at selected $V_{gate}$, when **(b)** $B//y$ and **(c)** $B//z$. All the measurements were carried out under $V_{SD}$ = 10 mV at 2 K.

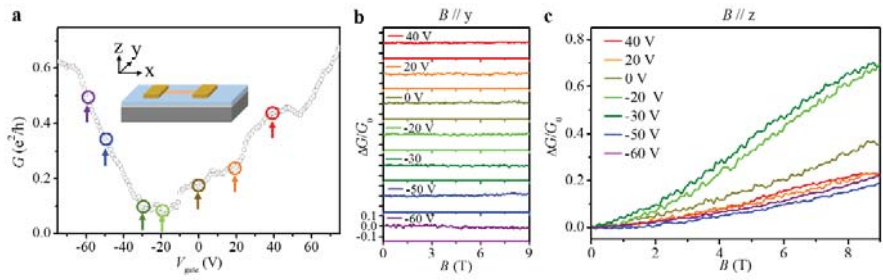

**Fig. S28. Magneto-transport in Type IV AGNR (sample #A85). (a)** Transfer characteristic for the 8.7 ± 0.5 nm AGNR; Normalized MC measured at different $V_{gate}$ indicated by the colored dots marked on the transfer curve shown in (a), when **(b)** $B//y$ and **(c)** $B//z$. All the measurements were under $V_{SD}$ = 10 mV at 2 K.

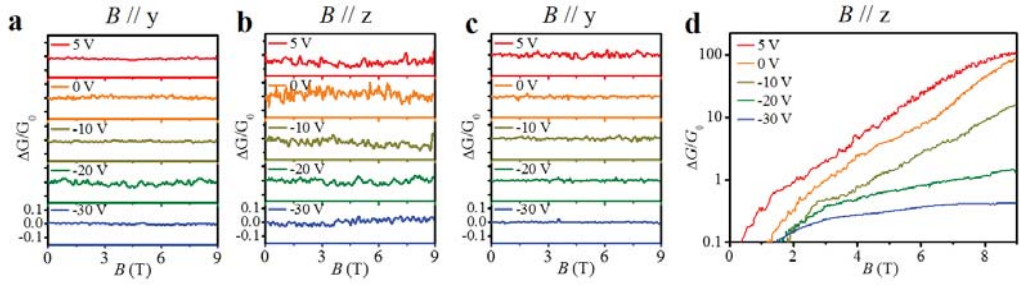

**Fig. S29. Magneto-electronic properties of GNR FETs under different $V_{gate}$.** (a-b) Normalized MC for a 5 nm-wide ZGNR (sample #Z143) measured at selected $V_{gate}$, (a) $B//y$ and (b) $B//z$. All the measurements were carried out at 4 K with $V_{SD}$ = 20 mV. (c-d) Normalized MC for a 4.8 nm-wide AGNR (sample #A187) measured at different $V_{gate}$, when (c) $B//y$ and (d) $B//z$. All the measurements were carried out at 2 K with $V_{SD}$ = 10 mV.

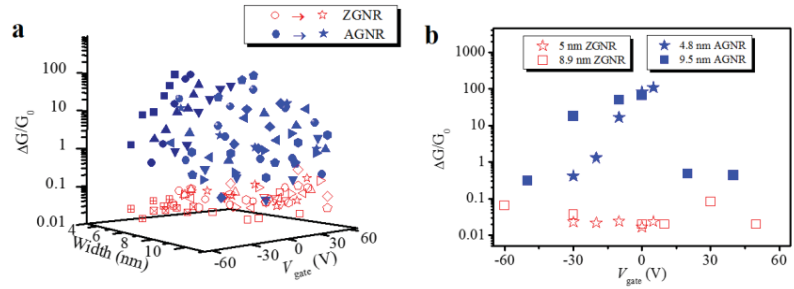

**Fig. S30. Distribution of MC of all GNRs measured. (a)** Normalized magneto-conductance as a function of $V_{gate}$ for GNR in different width subjected to a magnetic field of 9 T perpendicular to the substrate. The red open symbols represent the normalized MC for ZGNR samples and the blue solid symbols represent AGNR samples. **(b)** Value of normalized MC as a function of $V_{gate}$ for typical GNRs with some typical width subjected a magnetic field of 9 T perpendicular to the substrate for 5 nm-wide ZGNR (sample #Z143), 8.9 nm-wide ZGNR (sample #Z196), 4.8 nm-wide AGNR (sample #A187) and 9.5 nm-wide AGNR (sample #A135), respectively.

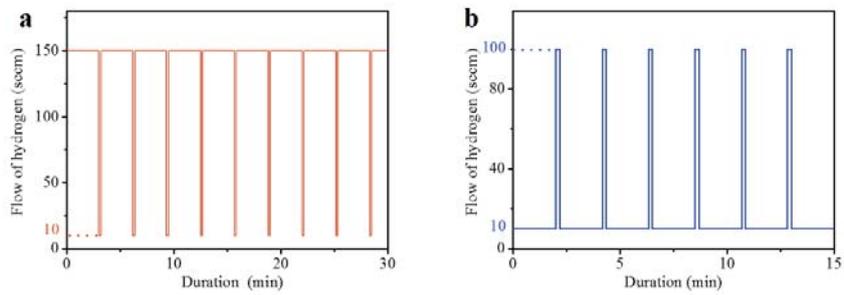

**Fig. S31. Schematic diagrams of typical processes for etching nanotrenches on *h*-BN to shorten the smooth segments. (a)** A typical etching process for ZZ-oriented trenches. The NiCl$_2$-coated samples were annealing at 1,200 °C for 30 min under a 10 s outage of hydrogen flow 150/10 sccm every 3 min. **(b)** A typical etching process for AC-oriented trenches. The H$_2$PtCl$_6$-coated samples were annealed at 1,300 °C for 15 min under a 10 s pulse of hydrogen flow 10/100 sccm every 2 min.

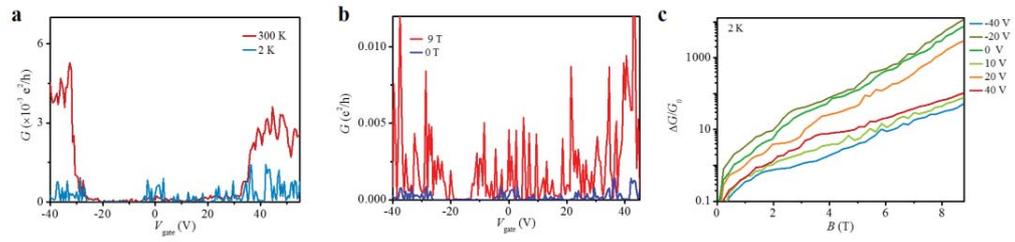

**Fig. S32. Electrical transport properties of a FET made from sub-5 nm wide ZGNR with short smooth segments.** Its channel length $l$ is ~240 nm. **(a)** $G$-$V_{gate}$ characteristics of the GNR device under 300 and 2 K. **(b)** $G$-$V_{gate}$ characteristics of the device subjected to different magnetic fields ($B$) at the temperature of 2 K. **(c)** Normalized magnetic conductance (MC) measured at different $V_{gate}$. All the measurements were under $V_{SD}$ = 20 mV.

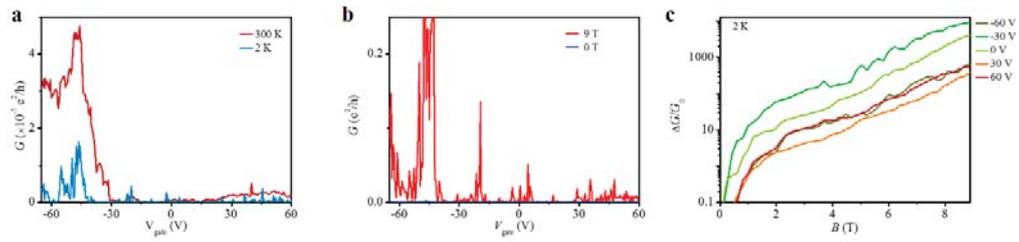

**Fig. S33. Electrical transport properties of a FET made from sub-5 nm AGNR with short smooth segments.**
**(a)** Transfer characteristics of the GNR device under 300 and 2 K. **(b)** Transfer curves for the device subjected to different $B$ at the temperature of 2 K. **(c)** Normalized MC measured at different $V_{gate}$. All the measurements were under $V_{SD}$ = 20 mV.

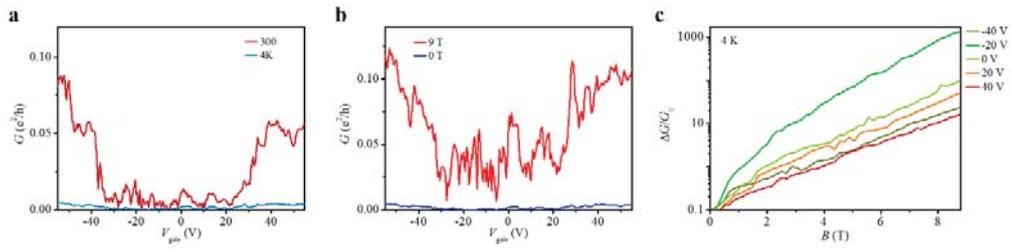

**Fig. S34. Electronic transport properties of a FET made from ~10 nm ZGNR with short segments in smooth edges. (a)** Transfer curves of the GNR device under 300 and 4 K. **(b)** Transfer curves for the device subjected to different $B$ at 4 K. **(c)** Normalized MC measured at different $V_{gate}$. All the measurements were carried out under $V_{SD}$ = 20 mV.

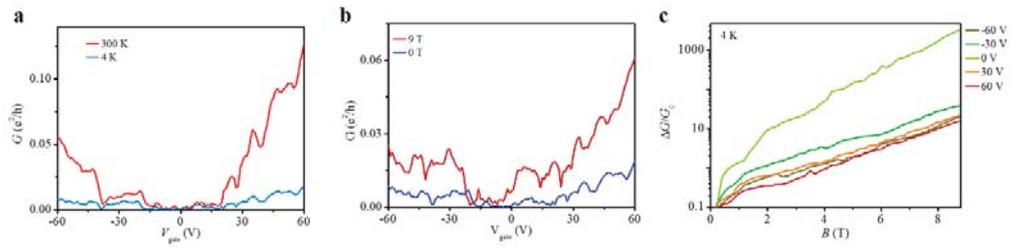

**Fig. S35. Electronic transport properties of a FET made from ~10 nm AGNR with short smooth segments.** **(a)** Transfer curves of the GNR device under 300 and 4 K. **(b)** Transfer curves for the device subjected to different $B$ at 4 K. **(c)** Normalized MC measured at different $V_{gate}$. All the measurements were under $V_{SD}$ = 20 mV.

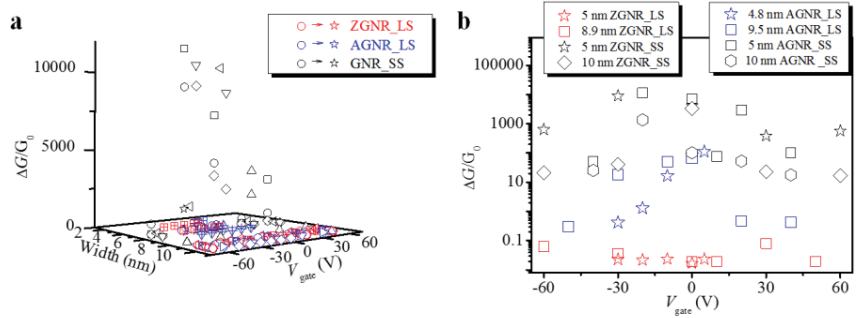

**Fig. S36. Distribution of MC of all GNRs measured. (a)** Normalized MC as a function of $V_{gate}$ for GNR in different widths subjected to a magnetic field of 9 T perpendicular to the substrate. The red open symbols represent the normalized MC for ZGNR with long smooth segments (ZGNR_LS) samples, the blue solid symbols represent AGNR with long smooth segments (AGNR_LS) samples and the black open symbols represent GNRs with short smooth segments (GNR_SS). **(b)** Value of normalized MC as a function of $V_{gate}$ for typical GNRs with some typical width subjected a magnetic field of 9 T perpendicular to the substrate for 5 nm-wide ZGNR_LS (sample #Z143), 8.9 nm-wide ZGNR_LS (sample #Z196), 5 nm-wide ZGNR_SS (sample #SSZ002), 10 nm-wide ZGNR_SS (sample #SSZ023), 4.8 nm-wide AGNR_LS (sample #A187) and 9.5 nm-wide AGNR_LS (sample #A135), 5 nm-wide AGNR_SS(sample #SSA012), and 10 nm-wide AGNR_SS(sample #SSA022), respectively.

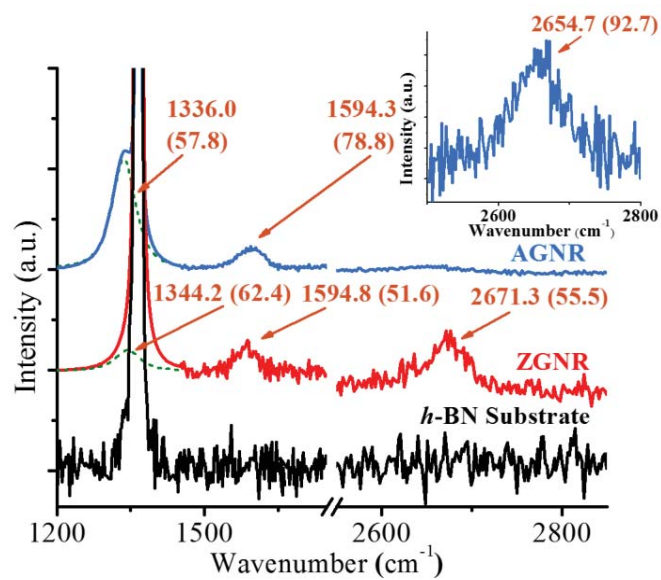

**Fig. S37. Raman scattering of a ZGNR and an AGNR on *h*-BN.** The Raman spectrum of *h*-BN is also shown beside those of the GNRs for comparison. The spectrum traces were normalized and shifted on the intensity axis for clarity. The inset shows a zoomed-in view of the 2D peak of AGNR. The fitted D peak (green) is extracted from the measured spectra because the position of the D-band of the GNRs is very close to that of a prominent Raman peak of *h*-BN. The full width at half-maximum (FWHM) for each peak is given in parentheses with the peak position.

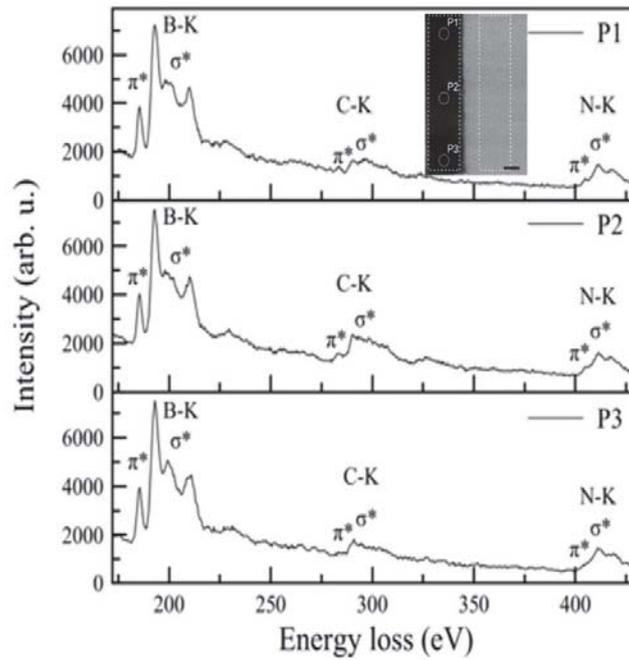

**Fig. S38.** EELS measurement on ZGNR embedded in *h*-BN. Inset shows the STEM-MAADF image of a ZGNR with the width ~5 nm, the scale bar is 3 nm. EELS spectrum recorded over the areas shown in the left dashed circles region (P1, P2, P3) in the inset.

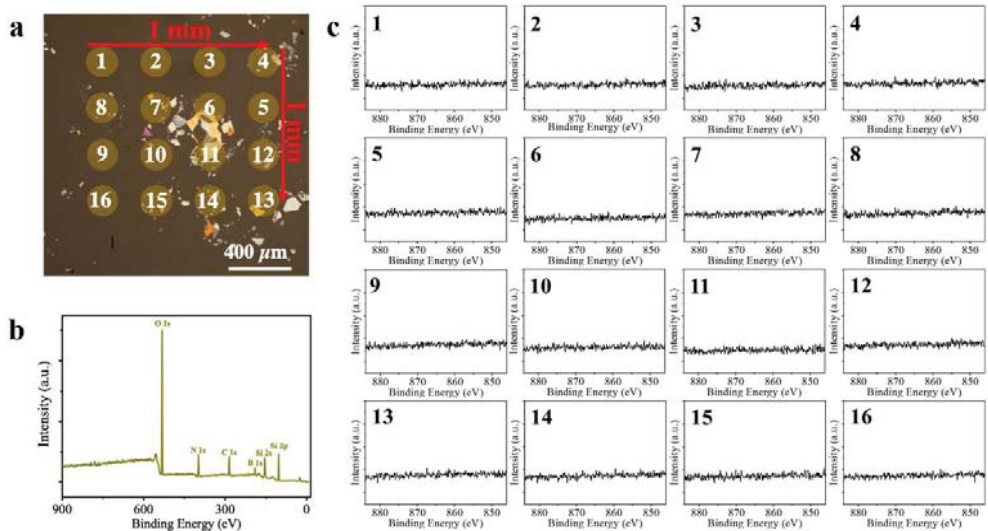

**Fig. S39. XPS analysis of embedded ZGNRs in *h*-BN flakes on quartz substrate after growth. (a)** Optical image of ZGNR/*h*-BN on quartz substrate. The XPS mapping area is 1 ×1 mm². **(b)** XPS survey spectrum of embedded ZGNR in *h*-BN flakes at position 11. The B 1s and N 1s signals are detectable. **(c)** Ni 2p spectra of embedded ZGNR in *h*-BN flakes randomly taken from 16 locations on the quartz substrate. The Ni signal is not detectable.

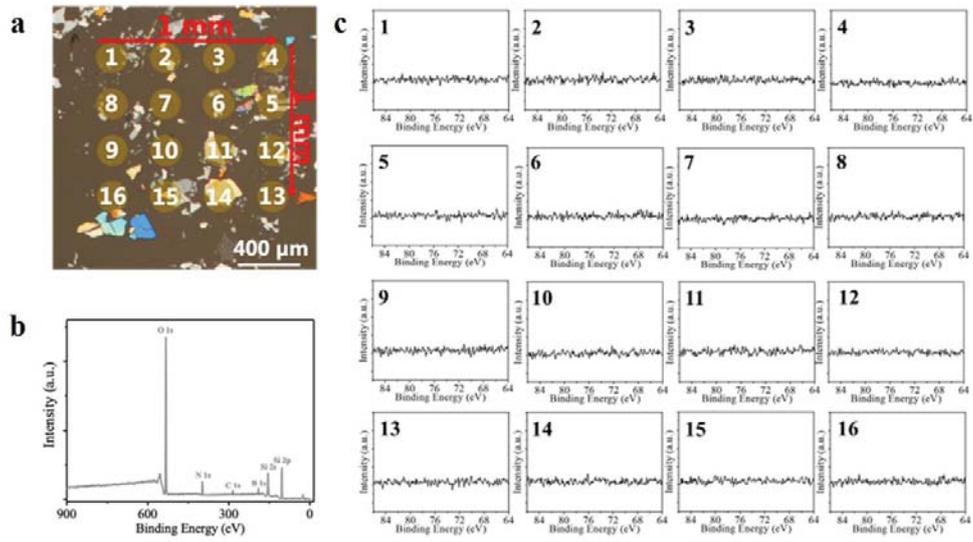

**Fig. S40. XPS analysis of embedded AGNRs in *h*-BN flakes on quartz substrate. (a)** Optical image of AGNR/*h*-BN on quartz substrate. **(b)** XPS survey spectrum of embedded AGNR in *h*-BN flakes at position 14. The B 1s and N 1s peaks are visible. **(c)** Pt 4f spectra of embedded AGNR in *h*-BN flakes taken from 16 locations on the quartz substrate. The Pt4f signal is not detected.

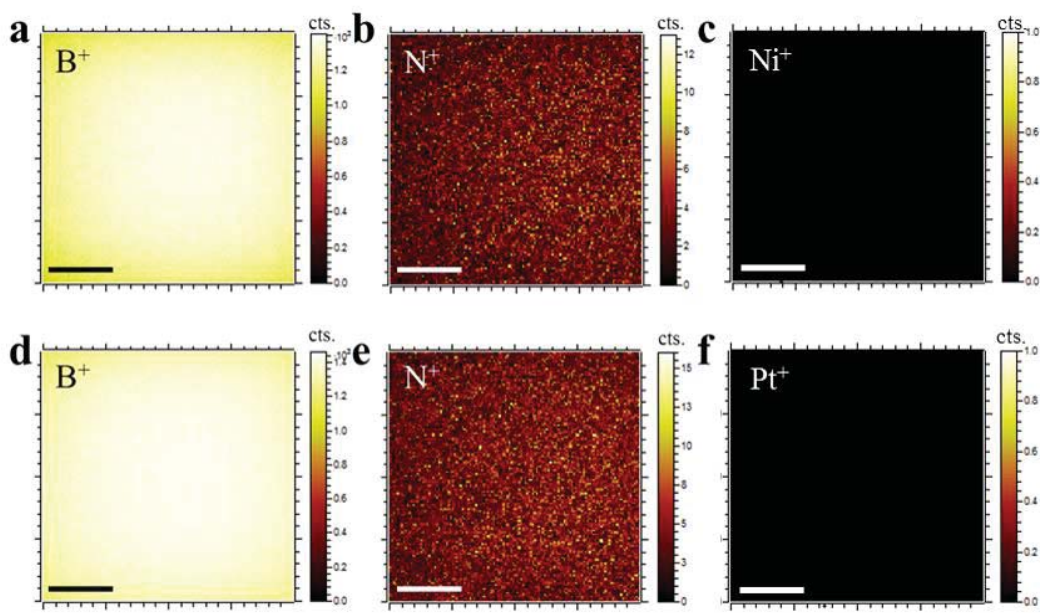

**Fig. S41. Planar distribution of the main elements on GNR sample. (a-c)** TOF-SIMS maps of B+, N+, and Ni+ secondary ions after the growth of ZGNR, respectively. The Ni+ signal corresponds to the nickel nanoparticles. **(d-f)** TOF-SIMS maps of B+, N+, and Pt+ secondary ions on an AGNR sample, respectively. The low Pt+ secondary ion signal indicates that the concentration of platinum nanoparticles after growth is too low to detect. Scale bars, 10 μm.

| Experiment | Atmosphere | Temperature (°C) | NiCl$_2$ solution concentration (g/L) | Etching duration (min) | Length (μm) | Width (nm) | Aspect ratio |
|---|---|---|---|---|---|---|---|
| #1 | Ar:H$_2$= 9:1 150 Pa | 1100 | 0.1 | 180 | N.A. | N.A. | N.A. |
| #2 | Ar:H$_2$= 9:1 150 Pa | 1150 | 0.1 | 60 | N.A. | N.A. | N.A. |
| #3 | Ar:H$_2$= 9:1 150 Pa | 1150 | 0.1 | 180 | ~0.5 | <10 | ~50 |
| #4 | Ar:H$_2$= 9:1 150 Pa | 1200 | 0.1 | 30 | ~1 | <5 | >200 |
| #5 | Ar:H$_2$= 9:1 150 Pa | 1200 | 0.1 | 60 | ~1.5 | <10 | >150 |
| #6 | Ar:H$_2$= 9:1 150 Pa | 1200 | 0.1 | 180 | ~2 | ~10 | ~200 |
| #7 | Ar:H$_2$= 9:1 150 Pa | 1250 | 0.1 | 15 | ~1 | ~5 | ~200 |
| #8 | Ar:H$_2$= 9:1 150 Pa | 1250 | 0.1 | 30 | ~1.5 | ~10 | ~150 |
| #9 | Ar:H$_2$= 9:1 150 Pa | 1300 | 0.1 | 15 | ~1.5 | ~15 | ~100 |
| #10 | Ar:H$_2$= 9:1 150 Pa | 1300 | 0.1 | 30 | ~1 | ~20 | ~50 |

**Table S1. Summary of nickel-particle-assisted etching experiments under different conditions.** Both the length and width of the *h*-BN trenches exhibit dependence on the experimental conditions, such as the pressure of the carrying gases, temperature, the concentration of the NiCl$_2$ solution, and the etching duration. The width of the trenches most strongly depends on the etching temperature and the duration. The dependent relationship makes sense as etching is normally based on a thermally activated reaction. In the experiment, we focus on fabrication of narrow trenches (sub-5 nm), so the etching atmosphere and NiCl$_2$ solution concentration were optimized and fixed. Compared to the length, the width of trenches is more sensitive to temperature and etching duration, while the evolution of the length and the width does not seem to follow the well-known Arrhenius law, most likely due to the involvement of additional etching mechanisms, for example, hydrogen-activated etching.

| Experiment | Atmosphere | Temperature (°C) | $H_2PtCl_6$ concentration (ml/L) | Etching duration (min) | Length (μm) | Width (nm) | Aspect ratio |
|---|---|---|---|---|---|---|---|
| #1 | Ar:$H_2$= 3:1 10 Pa | 1150 | 10 | 180 | N.A. | N.A. | N.A. |
| #2 | Ar:$H_2$= 3:1 10 Pa | 1200 | 10 | 60 | ~0.5 | ~10 | ~50 |
| #3 | Ar:$H_2$= 3:1 10 Pa | 1200 | 10 | 180 | ~1 | ~20 | ~50 |
| #4 | Ar:$H_2$= 3:1 10 Pa | 1250 | 10 | 30 | ~1 | ~10 | ~100 |
| #5 | Ar:$H_2$= 3:1 10 Pa | 1250 | 10 | 60 | ~1.5 | ~20 | ~75 |
| #6 | Ar:$H_2$= 3:1 10 Pa | 1250 | 10 | 180 | ~2.5 | ~30 | ~80 |
| #7 | Ar:$H_2$= 3:1 10 Pa | 1300 | 10 | 15 | ~2 | <5 | >400 |
| #8 | Ar:$H_2$= 3:1 10 Pa | 1300 | 10 | 30 | ~2.5 | ~15 | ~160 |
| #9 | Ar:$H_2$= 3:1 10 Pa | 1350 | 10 | 15 | ~1.5 | ~10 | ~150 |
| #10 | Ar:$H_2$= 3:1 10 Pa | 1350 | 10 | 30 | ~1 | ~20 | ~50 |

**Table S2. Summary of platinum-particle-assisted etching experiments under different condition.** Similar to the nickel-assisted-etching, both the length and width of the *h*-BN nano-trenches exhibit an obvious dependence on the experimental parameters. Compared to the ZZ oriented etching, the AC trenches are more sensitive to temperature and much easier to broaden.

# Supplementary Text

### DFT calculations for *h*-BN nano-trenches etching

We carried out additional calculations to understand the selective etching of *h*-BN by using Ni and Pt nanoparticles. Previous literature [1] reports that the selective cutting of graphene is along the armchair (AC) or zigzag (ZZ) direction which has strong interactions with the metal nanoparticle. Here, we calculated the interaction strength between *h*-BN edges and the Ni(111) and Pt(111) surfaces by using similar approaches.

To calculate the interfacial formation energy ($E_f$) between *h*-BN and the catalyst surface, we used *h*-BN nanoribbons with AC and ZZ edges attached to the surface of a three-layer transition metal (111) slab (the (111) surface is the most stable one) to represent the catalyst−BN interface. Another edge of the *h*-BN ribbon (NRs) is terminated by H atoms. The formation energy ($E_f$) of *h*-BN edge on the transition metal (TM) surface is defined as following:

$$E_f = E(H\text{-}BN\text{-}TM) - E(TM) - E(H\text{-}BN) + E_f(BN\text{-}vac),$$

where $E(H\text{-}BN\text{-}TM)$ and $E(TM)$ are the energies of the whole system and the TM(111) slab, respectively; $E(H\text{-}BN)$ is the energy of *h*-BN nanoribbon with one edge saturated by H atoms; L is the length of the TM(111)−BN interface in nanometers; $E_f(BN\text{-}vac)$ is the formation energy of *h*-BN edge in vacuum and it is calculated by the method used in the reference [2]. Based on the above equation, the $E_f$ of AC and ZZ on Ni(111) and Pt(111) surfaces were calculated, respectively. Considering that the termination atoms on ZZ edge can be B or N atom, the ZZ edge can be further classified as ZZ-B and ZZ-N edges. The formation energy of ZZ-B, ZZ-N and AC edges on Ni(111)/Pt(111) surface are 1.92/-1.07, -2.44/-2.42, -0.23/-1.03 eV/nm, respectively. Considering that the etching of *h*-BN along ZZ direction always results in both ZZ-B and ZZ-N edges simultaneously, we took the average formation energy of ZZ-B and ZZ-N as the formation energy of ZZ edge. Therefore, the formation energies of ZZ and AC on Ni(111)/Pt(111) surface are -0.26/-2.43 and -0.23/-1.03 eV/nm. On both metal surfaces, the formation energy of ZZ are lower than that of AC. The result means both Ni and Pt nanoparticle lead to the etching of ZZ edges.

We also tried the model of edges without any hydrogen atom. Unexpectedly, the formation energy of ZZ is lower than that of AC on both metal surfaces.

We also calculated the interaction strength between *h*-BN edges and the Pt(100) and (110) surfaces by using similar approaches. The large lattice mismatch includes some out-of-plane deformations into the calculation. The results from calculation also show that the formation energy of ZZ is lower than that of AC on Pt surfaces.

It seems that the etching mechanism of Pt nanoparticles in *h*-BN is very complex. Considering that the Ni and Pt nanoparticles have different melting point (Ni:1455°C, Pt:1768°C for bulk), we believe that the different etching behavior of Ni and Pt nanoparticles could be caused by the states of the nanoparticles (e.g. Ni nanoparticles may exist in a liquid phase while Pt nanoparticles could be in a solid state). Besides, the hydrogen pressure plays an important role in the etching. It is found that Ni nanoparticles lead to the etching of some AC oriented trenches in relatively high hydrogen pressure. Extensive investigations along the research direction are still undergoing in our research group. At this moment, we want to leave the question about etching mechanism open to experimental scientists and theorists.

**Energy profiles for the growth of GNR at *h*-BN edges**

Our experimental results demonstrated that the growth of graphene at the AC edge of *h*-BN is very fast without using any catalyst while the $SiH_4$ catalyst is required for the fast growth of graphene at the ZZ edges of *h*-BN. To understand the experimental observations, first principles DFT calculations was carried out. The edge configurations of *h*-BN edges can be zigzag boron (ZZB), zigzag nitrogen (ZZN), armchair (AC) edges and the corresponding H-passivated ones, depending on the experimental conditions. **Fig.** S5 shows the calculated phase diagram of ZZB, ZZN, AC and their H-passivated edges. It is clear that ZZB and ZZN can be passivated by H under very low hydrogen pressure, while the AC edge of *h*-BN can be stable even at a very high H pressure. Therefore, the pristine AC edge, H-passivated ZZB and ZZN edges should be the energetically favorable edges at very high temperature in the presence of moderate $H_2$ pressure. Following calculations about graphene growth will be carried out along these edges.

Based on the configurations of the three edges, we further calculated the energy barriers of graphene growth at the three edges. The activation barrier of graphene growth at the H-passivated ZZB edge is 0.96 eV (**Fig.** S6a), which is much higher than that (0.15 eV) at the AC edge (**Fig.** S8). However, the use of $SiH_4$ catalyst can reduce the growth barrier to only 0.01 eV (**Fig.** S6b), which greatly promote the growth rate of graphene at the H-passivated ZZB edge. On the other hand, the growth barrier of graphene at the H-passivated ZZN edge is also very high (1.96 eV), as shown in **Fig.** S7a. The use of $SiH_4$ catalyst can reduce the growth barrier to only 0.37 eV, as shown in **Fig.** S7b. **Fig.** S8 shows the energy profile of graphene growth through the addition of $C_2H_2$ into the AC edge. It can be seen that the energy barrier for graphene growth at the AC edge is very low (0.15 eV), demonstrating the growth of graphene at the AC edge is very fast even without the $SiH_4$ catalyst. The use of $SiH_4$ catalyst can further decrease the growth barrier, making the growth process to be barrierless. Overall, the growth of graphene at the AC edge of *h*-BN is very fast because of the low activation barrier, while graphene growth at the ZZB and ZZN edges are much slower and can be greatly improved by the using of $SiH_4$ catalyst.

# Calculation of capacitance and field effect mobility

Field effect transport of transistors made from GNRs grown in *h*-BN was measured. The gate voltage ($V_g$) dependence of the conductance at different temperatures is plotted in **Fig.** 3 and Supplementary **Fig.** S19-21. The carrier mobility and on-off ratio of each GNR transistor can be extracted from the plots.

According to a classical model, the capacitance is completely determined by the object's geometry and a dielectric constant of the medium. If the object's size shrinks to a nanometer scale, a finite density of state (DOS) which originates from the Pauli exclusion principle should be considered. Low-dimensional systems, having a small DOS, are not able to accumulate enough charge to completely screen the external field. In order to describe the effect of the electric field penetration in a finite DOS system, quantum capacitance should be taken into account. According to theoretical calculations [3], the quantum capacitance can be ignored in devices with a thick dielectric layer. As we are using 300 nm SiO$_2$ as dielectric layer in these experiments, only classical capacitance is taken into account.

Three-dimensional electrostatic simulation was used to calculate the effective capacitance ($C_{gs}$) via Fast Field Solvers Software (FFSS), which is available at http://www.fastfieldsolvers.com. The modeled structure includes a large back-gate plane of highly-doped-Si, a dielectric bi-layer with the same lateral dimension as the back-gate (300 nm thickness and dielectric constant $\varepsilon_0$=3.9, for the bottom SiO$_2$ layer and 10 nm thickness and $\varepsilon_0$ = 4 for the *h*-BN top-layer) and a graphene layer with the experimental dimensions lying ~0.35 nm above the dielectric layer and two metal fingers with the experimental dimensions of the contacts were used to represent them. ~1 nm grids were used for the GNR in a width of *w* (nm) in the simulation. The obtained capacitances are: $C_{gs}$ = 3.71 pF·m$^{-1}$ for a *w* = 5 nm (Supplementary **Fig.** S19 and S20) and $C_{gs}$ = 3.63 pF·m$^{-1}$ for a *w* = 4.8 nm (Supplementary **Fig.** S21). Based on the standard transistor model, the intrinsic carrier mobility is $\mu = \dfrac{\frac{dG}{dV_{gs}} \cdot L}{C_{gs}}$, where *L* represents the channel length of the GNRs; *G* represents the channel conductance measured. The electrical field mobility can then be derived for the GNR transistors at 300 K (shown in Supplementary **Fig.** S19-21).

## Simple two band (STB) model

For graphitic materials, a simple two band (STB) model [4-6] is always used to extract the band gap information from the temperature-dependence of the measured resistance. Based on this model, the densities of the electrons ($n$) and holes ($p$) are given by $n = C_n k_B T \ln(1+\exp(-\frac{E_C-E_F}{k_B T}))$ and $p = C_p k_B T \ln(1+\exp(-\frac{E_F-E_V}{k_B T}))$, respectively. Here, $E_F$, $E_C$, $E_V$ are the energies at the Fermi level, the bottom of the conduction band and the top of valance band, respectively, $k_B$ is the Boltzmann constant and $C_n$, $C_p$ are constants independent of temperature ($T$). Ignoring the contribution from static scattering centers, the mobility of the carriers can be expressed as $\mu_e = A_1 T^{-1}$ and $\mu_h = A_2 T^{-1}$, where $A_1$ and $A_2$ are constants depending on the strength of the electron and hole-phonon scattering in graphite. Since the resistivity is given by $\rho = (n\mu_e e + p\mu_h e)^{-1}$, the temperature dependence of the resistance can be expressed as:

$$R = \frac{P_1}{\ln(1+\exp(-\frac{E_C-E_F}{k_B T}))+P_2 \ln(1+\exp(-\frac{E_F-E_V}{k_B T}))} + R_{contact}$$

Using $E_C$-$E_F$, $E_F$-$E_V$, $P_1$, $P_2$ and $R_{contact}$ as the fitting parameters, this model fits well with our experimental data for most of the samples at different $V_{gate}$, as shown in supplementary **Fig.** S20. Based on this model, the energy gap obtained for the ZGNR sample (#Z143) with a width of ~5 nm is 436.2±28.1 meV. And the extracted band gaps are about 183.2±18.7 meV and 530.9±34.7 meV for sample #A39 and #A187, respectively. The model has ruled out a contribution from the contacts.

**Discussion on magneto-transport in GNR**

The GNRs embedded in *h*-BN provide an ideal experimental subject to investigate the origination of MC behavior in GNR. Imperfect edges are expected to produce significant scattering sources. Sources of disorder such as bulk vacancies, charges in the oxide, or structural deformations are also believed to alter the conductance, although the dominant scattering source remains debated.

In graphene, impurities scatters charge carriers in all possible direction, and then cause a slightly positive MC in a very low magnetic field. The phenomenon is named as weak localization [7,8]. If the spin-orbit interaction is sufficiently large, a quantum interference results in a negative MC behavior in graphene, a phenomenon referred weak anti-localization [9,10].

In GNRs, the positive MC results from a subtle interplay between the specific magnetic band structures and the field-induced reduction in impurity-driven backscattering. In the magnetic field we applied, the size effect of GNRs is dominated as the cyclotron length is greatly larger than the width of GNRs [11]. The charge transport due to the formation of cyclotron orbits originating from Dirac-Landau-level behavior cannot happen in the GNRs.

In narrow ZGNRs, a semiconducting anti-FM spin state theoretically can be excited to the metallic ferromagnetism (FM) state with a high enough magnetic field [12]. However, the semiconducting states in narrow ZGNRs can experimentally survive even at room temperature. Therefore, the magnetic field we applied here is not high enough to break the anti-FM exchange coupling in narrow ZGNR. In addition, the edge states in wide ZGNRs are conductive channels, and cannot be modified in the energy dispersion by external magnetic field. It may be the reason why we found that MC in the ZGNRs (for both Type I and II GNR) is not modified by the perpendicular magnetic field.

In AGNRs, tight-binding calculations have shown that the confinement bandgap of semiconducting AGNRs indeed shrinks continuously with increasing magnetic field [13]. The magnetic field dramatically modifies the energy dispersions and it changes the size of the bandgap, shifts the band-edge states, destroys the degeneracy of the energy bands, induces the semiconductor-metal transition and generates the partial flat bands. This is why the AGNRs exhibits more pronounced MC than ZGNRs does.

In addition, the edge roughness of GNRs and long-range Gaussian potential could cause a positive MC in the magnetic field [14]. The GNRs with more disordered edges or local potential variations have been observed with more pronounced MC [15,16].

It is noted that a relatively large positive MC is observed at low carrier density in AGNRs, and often become saturated at high magnetic field. The reason could be that more channels are available for conduction when higher gate-voltage is applied. The increase in conductance at higher charge density can be attributed to an increase in localization length, due to the increase in the number of occupied transverse channels [17]. In addition, the MC contribution could come from the so-called forward interference between two hopping sites [18]. In this condition, multiple conduction pathways between two points interfere to increase the probability of going from one site to other site, thus generating positive MC, this interference produces a positive MC that saturates at large magnetic field in GNRs.

Although the exact mechanism responsible for the observed difference of GNRs in their MC is not clear at this moment, our experimental findings demonstrate that the ZGNRs exhibit relatively small MC while AGNRs have higher MC. Further experiments well designed and theoretical studies will be carried out to understand the mechanism responsible for the MC behavior in GNRs.

As shown in all atomic resolution STEM images of GNRs, no metal atom is detected in the lattice of GNRs. In addition, all the magneto-transport results are reproducible in all the samples we measured. We believe that there is no influence in MC from the metal residues.

**Influence of edge roughness on magneto-transport in edge-specific GNRs**

Actually, we can reduce the length of segment with smooth edges in GNRs by modifying the template of $h$-BN nanotrenches. Both ZZ-oriented and AC-oriented GNRs with short smooth segments exhibit similar behavior in their transfer curves and magneto-conductance, which are obviously different from those GNRs with long smooth segments. Based on the results of transport measurement, it is believed that the edge-specific GNRs with long smooth segments embedded in $h$-BN have been successfully fabricated in $h$-BN substrates. The details about the experiments are shown below.

We intentionally changed the partial pressure of hydrogen during etching to obtain $h$-BN nanotrenches with short smooth segments. The flow of $H_2$ was reduced from 150 to 10 sccm every 3 min for 10 seconds during ZZ-oriented etching, while the flow of $H_2$ was switched from 10 to 100 sccm every 2 min for 10 seconds during AC-oriented etching, as shown in **Fig.** S31. The unstable etching environment would produce the trenches with short smooth segments. With these templates, GNRs with short smooth segments can be fabricated.

For sub-5 nm-wide ZZ-oriented GNR samples with short smooth segments, we investigate their electrical properties at different temperatures and magnetic fields. The results are shown in **Fig.** S32. Figure S32a shows conductance ($G$) as a function of $V_{gate}$ at room temperature and 2 K of the typical ZZ-oriented GNR FET with short smooth segments. Its channel length $l$ is ~240 nm and the width $w$ is ~5 nm. The ZZ-oriented GNR device shows about 3 orders lower conductivity of on-state and much lower mobility than those ZGNRs with long smooth segments. The on-state conductance is ~$5\times10^{-3}$ $e^2/h$ at 300 K. And the GNR shows suppressed conductance over a wide range (from -65 to 50 V) in the $G$-$V_{gate}$ characteristics at 2 K, with some fluctuations observed inside. Such effects are always attributed to transport gap resulting from edge roughness, similar behavior has been observed in lithographic GNRs [19] and graphene nano-constrictions [20]. Theoretical calculations [21] have shown that the density of states of GNRs with rough edge are dominated by localized states, and the charge transport could be through hopping between these localized states at low temperature.

Figure S32b shows $G$ as a function of $V_{gate}$ under different magnetic fields perpendicular to the substrate at 2 K. Although the conductance still shows fluctuations in the whole gate range, the GNR shows a very large positive MC near the minimum conductance. Upon applying a magnetic field, the overall conductance increases dramatically, with a reduced transport gap (from -30 to 35 V). These results suggest that the transport scattering decreases with increasing magnetic field.

Figure S32c shows the magneto-transport properties of the ZZ-oriented GNR device, a huge positive MC (more than a factor of 10,000) is found at 2 K. It is clear that the conductance is essentially switched on from an off-state. And it is also found that all the MCs are positive at all $V_{gate}$ and there is no trend for saturation of the MCs. The large MC normally appears near the suppressed transport gap regime.

Similarly, we also investigated the electrical transport of AC-oriented GNRs with short smooth segments. The results are shown in **Fig.** S33. Figure S33 shows the electrical transport properties of an AGNR with short smooth segments. Its channel length $l$ is ~220 nm and a width $w$ of ~5 nm. As **Fig.** S33a shows, the on-state conductance is ~0.005 $e^2/h$ and the transport gap is more than 60

V even at 300 K. Figure S33b shows that the conductance of the AGNR can also be tuned by the magnetic field. And the value of MC is also very large at the suppressed transport gap regime. As **Fig.** S33c shows, the MCs are also positive and large at all the gate biases. And a huge positive MC of ~10,000 is occurred at $V_{gate}$ = -30 V at 9 T. All the transport behavior is quite similar to the ZZ-oriented GNR with short smooth segments.

Both ZGNRs and AGNRs with short smooth segments include much more mixed edges and then have higher edge roughness, the GNR would like to behave as a series number of quantum dots with large numbers of localized sites, which may cause a low conductivity. The edge roughness induced back-scattering, in terms of strong localization, may contribute to the observed conductance suppression and which is more sensitive to the magnetic field. And this is the reason why the sub-5 nm-wide ZGNRs and AGNRs with short smooth segments show similar transport behavior.

In addition, we also studied the electronic properties for wider GNRs $w \sim$ 10 nm. The results are shown in **Fig.** S34 and S35.

As shown in **Fig.** S34, the on-state conductance of the ZGNR device (channel length $l$ of ~245 nm) is ~ 0.09 $e^2/h$, which is higher than the sub-5 nm-wide ZGNR device. The transport gap is about 40 V at 4 K. Figure S34b shows the transfer curves for the device subjected to different $B$ at 4 K. It also can be found that the $B$ can tune the conductance and the factor is much smaller when the device is gated far away from the transport gap. Figure S34c shows the MCs at different $V_{gate}$ at 4 K. The relatively large positive MC ~2,000 is observed when the $B$ is at 9 T and the gate voltage is set to 0 V.

Comparison to the ~10 nm-wide ZGNR with short smooth segments, the 10 nm-wide ZGNR with long smooth segments show a much higher on state conductance. Significantly, the weak metallic conductance peak which is related to the zigzag edge is not detected which means the roughness of ZGNR edge increases. Besides, the MCs in ZGNRs are 2-3 orders higher than those in ZGNRs with long smooth segments. Apparently, in this case, the huge MC behavior attributes to increase of edge roughness.

Similarly, we also carried out the transport measurement of low quality AGNR FET with w ~ 10 nm. The results are shown in **Fig.** S35. The channel length $l$ of ~228 nm. As shown in **Fig.** S35a, the on-state conductance is a ~0.2 $e^2/h$ at 300 K and ~0.015 $e^2/h$ at 4 K. And the transport gap at 4 K is about 30 V. The magnet filed can also tuned the conductance at 4 K (**Fig.** S35b). Figure S35c shows that a relatively large positive MCs is obtained at all $V_{gate}$.

We plot all values of MC obtained under $B_{//z}$ = 9 T at different $V_{gate}$ for all GNR samples in **Fig.** S36. As shown in **Fig.** S36, MCs in ZGNRs with short smooth segments become very large, and the values are quite similar to those in AGNRs with short smooth segments. Some of them are higher than 10E5, as shown in **Fig.** S36a. The large MC mainly comes from scattering of rough edges to charge carriers in both ZGNRs and AGNRs with short smooth segments. Both the ZGNRs and AGNRs with short smooth segments show much larger positive MCs than GNRs with longer smooth edges, the results are believed to be related to edge roughness.

In summary, both ZGNR and AGNR with short smooth segments show much lower on state conductance and wider transport gap than those with long smooth segments. For ZGNR with short smooth segments, both the sub-5 nm-wide and 10 nm-wide ZGNRs show 2-3 orders lower on-state conductance and 2 orders higher MC than those with long smooth segments when the magnetic field is applied to 9 T. Besides, a pronounced conductance peak is absent in 10 nm-wide ZGNRs with short smooth segments while that is always observed in those ZGNRs with long smooth segments. For AGNR, on-state conductance in AGNRs with short smooth segments becomes 2-3 orders lower than those with long smooth segments, and their MCs increase 2-3 orders higher than those with long smooth segments. It is also found that the MCs of ~10 nm-wide AGNR don't show any trend of saturation which happens in AGNRs with long smooth segments.

# Raman Spectra of GNRs

Raman measurements were carried out to investigate the structural and electronic properties of the ultra-narrow AGNR and ZGNR on $h$-BN. All the data were collected using an exciting laser of 532 nm. **Fig.** S37 shows the Raman spectra. In all spectra shown in **Fig.** S37, a prominent sharp peak appeared at ~1,366 cm$^{-1}$, which was attributed to the Raman-active LO phonon of $h$-BN [22]. For the spectra of the GNRs, the G-, D- and 2D-bands were fitted with a single-Lorentzian line shape. In the spectrum of the AGNR (azure line), the D-band appeared at 1336.0 cm$^{-1}$, and G-band at 1594.3 cm$^{-1}$, and the inset shows an enlarged 2D-band of the AGNR(~2,654.7 cm$^{-1}$). In the spectrum of ZGNR (red line), the D band appears at 1,344.2 cm$^{-1}$, the G-peak at 1,594.8 cm$^{-1}$ and the 2D-peak at ~2,671.3 cm$^{-1}$. Raman spectra confirm that the narrow GNRs consist of sp$^2$ carbon. In addition, the observed broadening of the full width at half-maximum for the Raman G- and 2D-bands in both AGNR and ZGNR may be due to strain variations at the nanometre-scale [23]. It is obvious that the D peak in AGNR exhibits much stronger intensity than that of ZGNR because the defect-assisted double resonant inter-valley scattering process only occurs at armchair edges and is forbidden at the zigzag edge [24]. The tiny D band in ZGNR may come from imperfections in the zigzag edges. In most cases, a ZGNR is comprised of series of perfect zigzag segments. The average length of the straight segments can be estimated by measuring the mean free path of charge carriers from transport measurement.